\documentclass[letterpaper,prb,aps,floatfix,reprint,notitlepage,nofootinbib, 10pt, reprint, twocolumn, superscriptaddress]{revtex4-2}
\usepackage{graphicx} 
\usepackage{amsfonts}
\usepackage{amsthm}
\usepackage{mathtools}
\usepackage[dvipsnames]{xcolor}
\usepackage[colorlinks=true]{hyperref}
\hypersetup{
    pdftitle={Toom Ratchets},
    citecolor=Green,
    linkcolor=blue,
}
\usepackage{thmtools}
\usepackage{thm-restate}
\usepackage[normalem]{ulem}
\usepackage{mdframed}

\newtheorem{prototheorem}{Theorem}
\newtheorem{protoresult}{Informal Result}
\newtheorem{protodef}{Definition}

\newenvironment{theorem}
  {\begin{mdframed}[backgroundcolor=gray!15, linewidth=0pt]
   \begin{prototheorem}}
  {\end{prototheorem}
   \end{mdframed}}

\newenvironment{result}
  {\begin{mdframed}[backgroundcolor=gray!15, linewidth=0pt]
   \begin{protoresult}}
  {\end{protoresult}
   \end{mdframed}}

\newenvironment{definition}
  {\begin{mdframed}[backgroundcolor=gray!15, linewidth=0pt]
   \begin{protodef}}
  {\end{protodef}
   \end{mdframed}}

\DeclareMathOperator\sgn{sgn}
\newcommand{\mca}{{\mathcal A}}
\newcommand{\mcr}{{\mathcal R}}
\newcommand{\mcp}{{\mathcal P}}
\newcommand{\s}{\sigma}
\newcommand{\size}{r}
\newcommand{\async}{\mathrm{async}}
\newcommand{\sync}{\mathrm{sync}}

\usepackage{tikz}
\usetikzlibrary{matrix, positioning}

\begin{document}
	
\begin{abstract}
    Active systems can exhibit a broad range of phenomena forbidden in equilibrium. Their dynamics are often specified by abstract local update rules, and it is generally unclear when the same behavior can arise from physically natural driving. Here we show that two simple driving mechanisms can universally simulate any local active dynamics in spin systems. The first is the familiar setting of a time-periodic Hamiltonian coupled to a cold bath, which we call a \emph{many-body Brownian pump}. As a second mechanism, we promote the Brownian ratchet, traditionally a mechanism for transport, to a \emph{many-body Brownian ratchet}: a static Hamiltonian coupled to a hot bath and a cold bath, where the resulting steady heat current can be harnessed not only to drive transport but also to generate local active dynamics. Using probabilistic cellular automata as an explicit model, we prove that for any continuous-time (or discrete-time) local active dynamics, there is always a many-body Brownian ratchet (or pump) that approximates the dynamics, up to noise that can be made arbitrarily weak by tuning energy scales and other parameters. As a concrete demonstration, we construct a simple ferromagnetic Ising ratchet on a bilayer lattice. When the two layers are coupled to baths at different temperatures, this model serves as a robust classical memory even under a symmetry-breaking field, something impossible in equilibrium. More broadly, our work shows that ratchets can use steady heat currents to autonomously generate and stabilize novel collective behavior, realizing a new static setting for nonequilibrium many-body dynamics.
\end{abstract}

\title{Brownian ratchets and pumps universally simulate many-body active dynamics}
\author{Charles Stahl}
\affiliation{Department of Physics, Stanford University, Stanford, California 94305, USA}
\author{Ethan Lake}
\affiliation{Department of Physics, University of California Berkeley, Berkeley, California 94720, USA}
\author{Vedika Khemani}
\affiliation{Department of Physics, Stanford University, Stanford, California 94305, USA}

\maketitle

\section{Introduction}

Active dynamics are everywhere, completing tasks that are impossible in equilibrium. Birds flock, breaking symmetry in a way forbidden in equilibrium~\cite{Toner1995, Toner1998, Toner2005}. Molecular motors walk along filaments, converting chemical energy into directed motion \cite{bustamante2001physics}. Elsewhere, mechanisms ranging from vision cones~\cite{Loos2023} to
nonreciprocal~\cite{You2020, FruchartNonreciprocity, Nadolny2025} and constrained dynamics~\cite{Shi2025} all provide examples of interesting nonequilibrium phenomena.
What unites these systems is that they are all driven: they consume energy, produce entropy, and break detailed balance. They also carry steady-state currents---of energy or information---and these currents can be harnessed to do useful work. 

\begin{figure}
	\centering
	\includegraphics[]{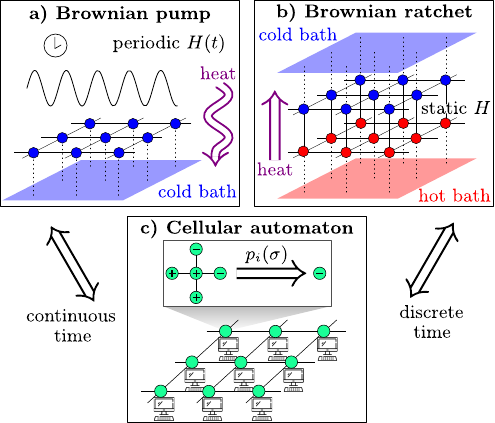}
	\caption{
    We prove that (a) time-dependent systems with dissipation (pumps) and (b) systems coupled to baths at different temperatures (ratchets) can simulate (c) cellular automaton dynamics, in continuous and discrete time, respectively.
    }
	\label{fig:paradigms}
\end{figure}

One standard way to drive a many-body system out of equilibrium is to make its Hamiltonian periodic in time: the time dependence injects work and can generate phenomena with no equilibrium analogue, from time crystals~\cite{KhemaniTC, ElseTC, Machado2023, SidToom} to synchronization~\cite{Acebron2005KuramotoReview, Gupta2014KuramotoReview}. In closed systems the drive generically causes heating, while in open systems a cold bath can absorb entropy and stabilize the driven dynamics, giving the \emph{Brownian pump} setting studied below and illustrated in Fig.~\ref{fig:paradigms}(a). Here we propose a complementary route to active many-body dynamics: the many-body \emph{Brownian ratchet}, which is a \emph{static} Hamiltonian coupled to two baths at different temperatures, illustrated in Fig.~\ref{fig:paradigms}(b). A heat current flowing from hot to cold through the system can do work, like in a heat engine. 
Brownian ratchets have long served as mechanisms for rectification and transport~\cite{Reimann2002BrownianMotors, Hanggi2009}; here we introduce them instead as a new playground for nonequilibrium many-body physics, in which steady heat currents power active dynamics and stabilize collective behavior that is impossible in equilibrium. In particular, we show that Brownian ratchets and pumps can universally simulate \emph{any} local active dynamics.

To make this statement precise, we need a general language for local active dynamics. Active systems are often described in terms of interactive dynamics, whereby a large number of energy-consuming entities (birds, bacteria, etc.) independently acquire information about their neighborhood and then use this information to update their positions or internal states. A canonical abstraction of such dynamics, illustrated in Fig.~\ref{fig:paradigms}(c), is a cellular automaton~\cite{vonNeumann1966, wolfram1983statistical}, which defines update rules for a system of discrete degrees of freedom based only on their local neighborhood. More generally, probabilistic cellular automata (PCAs) allow the local update to be stochastic. This rule-based language can realize computation, pattern formation, and robust information storage~\cite{vonNeumann1966, wolfram1983statistical, gardner1970life, Toom1980, Gacs2001}. In particular, Toom's rule~\cite{Toom1980} and related noisy cellular automata~\cite{Gacs2001, pajouheshgar2025exploring, lake2025squeezing} preserve information even under a bias favoring one state over the other, something impossible for an equilibrium system with a local order parameter. Thus, PCAs give sharp examples of local dynamics that evade equilibrium constraints.

Our rigorous results are theorems that show that Brownian pumps and ratchets are general enough to realize this entire class of local active dynamics. Brownian ratchets can approximate arbitrarily well the dynamics of any continuous-time PCA, meaning a local Markov process in which sites update asynchronously at random times. Brownian pumps can approximate arbitrarily well the dynamics of any discrete-time PCA, meaning a process in which all sites update synchronously in parallel. In this sense, Brownian ratchets and pumps \emph{simulate} abstract active dynamics in the spirit of Feynman~\cite{Turing1937, vonNeumann1966, Feynman1982, Gacs2001, cotler2025self}: one system reproduces the dynamics of another, while exposing a different physical structure or resource.

The physical value of these theorems is that they translate rule-based updates into the language of conventional condensed-matter driving, identifying simple thermodynamic resources for local active dynamics.  In a PCA, the nonequilibrium nature of the drive is hidden in the transition probabilities. In conventional condensed-matter settings, by contrast, the drive is packaged into a small number of external controls: alternating-current (AC) susceptibility or microwave spectroscopy experiments couple the system to a global time-dependent drive, while direct-current (DC) transport or Hall setups couple the system to reservoirs at different potentials or temperatures~\cite{kubo1991statistical, zwanzig2001nonequilibrium}. Brownian pumps and Brownian ratchets are the corresponding AC and DC routes to local active dynamics: a pump oscillates, while a ratchet maintains a steady temperature difference. The dynamics are generated autonomously by Hamiltonians, temperature gradients, heat currents, and dissipation, with no site-resolved measurement or feedback. These are natural variables of nonequilibrium statistical mechanics, making the resources that power the dynamics explicit, and allowing the machinery of nonequilibrium many-body physics to be brought to bear on local active dynamics.

For all simulation approaches, important considerations include overheads and error rates. In our constructions of Brownian ratchets and pumps, we only require a small overhead in space: 
to implement arbitrary active dynamics on $N$ degrees of freedom, our constructions use $cN$ degrees of freedom, for a small constant $c$. 
Because our constructions harness thermal fluctuations to drive the dynamics, they inherently implement noisy PCAs~\cite{Grinstein1985, GeorgesLeDoussal1989PCA, Lebowitz1990, Gacs2024Toom}.
They are therefore most ideally suited for noise-robust PCAs, in contrast to fine-tuned PCAs such as Conway's Game of Life~\cite{gardner1970life}, whose behavior becomes trivial under the addition of any nonzero amount of noise.
We show that the noise amplitude can be made arbitrarily small by tuning system parameters such as energy scales and temperatures appropriately. 

One particularly interesting class of noise-robust active systems is {\it many-body memories}: systems which autonomously and robustly store information even in the presence of generic noise \cite{Toom1980,Tsirelson,Gacs2001}.
The 2D Ising model, the paradigmatic classical memory, acts as a memory at low temperature by remembering its initial magnetization. The memory is fragile, however, because any external symmetry-breaking field pushes the system into the favored state in finite time (more generally, the Gibbs phase rule says that, in any equilibrium system with a local order parameter, spontaneous symmetry breaking is unstable to a symmetry-breaking field~\cite{Landau}).
On the other hand, certain PCAs that break detailed balance, like Toom's rule~\cite{Toom1980} and other more recent examples \cite{Gacs2001,pajouheshgar2025exploring,lake2025squeezing}, preserve information indefinitely even under a symmetry-breaking bias favoring one state over the other. 

As a concrete exploration of our  universal simulation results, we construct a particularly simple Brownian ratchet inspired by Toom's rule, in the form of a bilayer Ising model with ferromagnetic couplings, with one layer coupled to a hot bath and the other to a cold bath.  Despite its simplicity, this model preserves a memory even under a symmetry-breaking field, which as mentioned above is impossible in equilibrium. When the temperatures are the same, it reverts to an ordinary Ising ferromagnet with no robustness: the maintained temperature difference powers the error correction. We also provide a pump realization of Toom's rule, generalizing a previous result~\cite{Machado2023} from  oscillators to spins.

Developing dynamics which can protect {\it quantum} information in regimes where no equilibrium mechanism exists is also a subject of longstanding interest~\cite{Aharonov2008, Dennis2002, harrington2004analysis, Breuckmann2017LocalDecoders, Balasubramanian2024, LakeDecoder, lake2025local, Duennweber2026}.
As will be discussed in future work, our constructions extend naturally to quantum systems. Our results show that a local decoder for a quantum error-correcting code, once expressed as measurement-and-feedback PCA dynamics, can be translated into the thermal dynamics of a fixed Hamiltonian coupled to two heat baths. The decoder runs autonomously,
powered by a temperature gradient, with no explicit measurement or feedback required. This is distinct from prior work on engineered dissipation~\cite{Pastawski2011, Terhal2015, Dengis2014, Fujii2014, Herold2015, Hong2025} and simulated cooling~\cite{Seetharam2015, Scarlatella2019, Diermann2019, Naseem2021, Wen2021, Veness2023, Mori2023}, which use driving to perform tasks that are in principle achievable by cooling to equilibrium. In this work, we use driving for tasks that are \emph{impossible} in equilibrium.

The paper is organized as follows. In Sec.~\ref{sec:warmup} we warm up by constructing systems with biased diffusion driven by a temperature gradient or a periodic drive. Section~\ref{sec:summary} summarizes the main results of the paper. In Sec.~\ref{sec:ratchet} we construct the Brownian Toom ratchet, a simple nonequilibrium Ising model that is stable to a symmetry-breaking field. Section~\ref{sec:generic} gives a generic ratchet construction for arbitrary PCAs. Then, Sec.~\ref{sec:pumps} gives the analogous pump construction. Finally, we discuss quantum extensions and open questions in Sec.~\ref{sec:discussion}. 

\section{Warm up: Biased diffusion} \label{sec:warmup}

As a warm up, we consider biased diffusion of a single particle on a ring in 1D. This is the simplest possible nontrivial active dynamics, because it has a steady-state current, which is forbidden in equilibrium. We consider three mechanisms for realizing biased motion: interactive dynamics with a fixed abstract update rule, Brownian ratchet dynamics with a static Hamiltonian coupled to two baths at different temperatures, and Brownian pump dynamics with a time-dependent Hamiltonian coupled to a bath. As described in the introduction, both autonomous constructions require driving and dissipation. 

\textbf{The interactive dynamics} of biased diffusion consist of a particle at position $x(t)$ with an update rule 
\begin{equation}
	x(t+1) = \begin{cases*}
		x(t) + 1 & with probability $p$ \\
		x(t) - 1 & with probability $q$ \\
		x(t) & with probability $1-p-q$,
	\end{cases*}
\end{equation}
with $x(t)$ taken mod $L$.
The result is a drift velocity $v = (p-q)/(p+q)$ and a diffusion constant $D = ( (p + q) - (q - p)^2 )/2$. As long as the drift velocity is nonzero, the particle will continue to move around the ring even in the steady state.

These dynamics cannot be realized as the dynamics of a thermal system. Any system that obeys detailed balance has to have the probability of hopping left and right ($p=q$) and therefore no drift velocity ($v=0$). Thus, the equilibrium steady state cannot have any currents.

\begin{figure}
	\centering
	\includegraphics[]{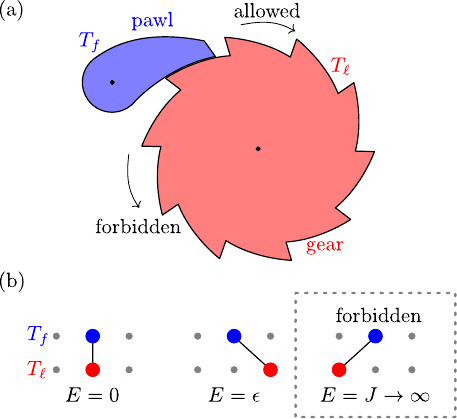}
	\caption{(a) A Brownian ratchet. If the gear is hotter than the pawl, then Brownian motion will apply random forces clockwise and counterclockwise, but the pawl will preferentially allow the gear to rotate clockwise. The result will is clockwise motion. (b) A Brownian ratchet on a lattice. In the limit $T_f \ll \epsilon \ll T_\ell$ the leader spin (bottom) can either be on the same site or one ahead of the follower spin (top), but the follower spin always wants to be on the same site as the leader.}
	\label{fig:ratchet}
\end{figure}

\textbf{The Brownian ratchet} is a time-independent Hamiltonian that is driven by virtue of different parts of the system being coupled to heat baths at different temperatures. This type of system acts as an autonomous heat engine, pulling useful work out of a temperature difference.  The classic ratchet consists of two components, a gear and a pawl (see Fig.~\ref{fig:ratchet}). The gear is connected to a hot bath, so that it is randomly driven by Brownian motion. The pawl only lets the gear spin in one direction. As long as the gear stays hot and the pawl stays cold, the ratchet acts as a heat engine, extracting useful work from the temperature difference.

To apply the Brownian ratchet principle to the particle on a ring, we consider a single leader particle $x_\ell$ (the gear) and a single follower particle $x_f$ (the pawl). We want to ensure that the follower particle always updates to match the leader spin, so $x_f=x_\ell$ should be the ground state of the system. We also want thermal fluctuations to allow the leader to walk ahead of the follower, so $x_\ell = x_f+1$ should have an energy that is small compared to the leader temperature. However, we don't want the leader to end up behind the follower, so $x_\ell=x_f-1$ should have a large energy penalty. Similarly, we want to strongly penalize configurations where the leader and follower are more than one site apart.

Packaging this all together, the Hamiltonian is
\begin{equation}
	H = \begin{cases}
		0, & x_\ell = x_f \\
		\epsilon, & x_\ell = x_f+1\\
		J, & \textrm{otherwise,}
	\end{cases}
\end{equation}
with $0 < \epsilon < J$. We couple $x_\ell$ to a bath at temperature $T_\ell$ (inverse temperature $\beta_\ell$) and $x_f$ to a bath at temperature $T_f$ (inverse temperature $\beta_f$).  The leader performs Glauber dynamics at temperature $T_\ell$, meaning it accepts updates with probability
\begin{equation}
p_\text{Glauber} = \frac{1}{1+e^{\Delta H / T_\ell}},
\end{equation}
at a clock rate $\Gamma_\ell$.
Similarly, the follower performs Glauber dynamics at temperature $T_f$ and clock rate $\Gamma_f$.\footnote{The choice of Glauber dynamics is somewhat arbitrary---other choices that obey detailed balance with respect to $H$ (like Metropolis updates) would work equally well.}

By clock rate, we mean the rate at which we propose updates to each spin.
In equilibrium dynamics, the clock rates are usually left unspecified, or set to all be equal. They have no effect on the steady state because every degree of freedom is locally in equilibrium. In nonequilibrium dynamics the clock rates do matter and can change the steady state. Throughout the paper, we will find the limit $\Gamma_f \gg \Gamma_\ell$ useful for analytical control, but we will perform our numerics at $\Gamma_f = \Gamma_\ell$.

The dynamics are particularly simple in the limit
\begin{equation}
T_f \ll \epsilon \ll T_\ell \ll J, \quad \Gamma_f \gg \Gamma_\ell,
\end{equation}
and on time scales much less than $e^{\beta_\ell J}$ so that we only have to consider the configurations where the leader is on the same site as or one site ahead of the follower (one can think of the leader and follower as constituting a diatomic molecule with two internal states). 
Because the follower updates very fast and $T_f\ll\epsilon$, it is nearly always in the ground state $x_f=x_\ell$. When the leader does update, it is nearly just as likely to step forward to $x_f+1$ or stay at $x_f$. The result is that the leader is the gear, taking occasional steps forward, and the follower is the pawl, nearly immediately stepping forward to prevent the leader from stepping back. The resulting random walk is nearly perfectly biased, in the sense that both particles only ever move forward, and happens in continuous time, in the sense that they do not take steps at any regular interval. 

If we keep the limit $T_f \ll \epsilon \ll T_\ell \ll J$ but set $\Gamma_f = \Gamma_\ell$, the dynamics remain nearly perfectly biased. When the follower updates, it is much more likely to update to $x_\ell$. When the leader updates, it is nearly just as likely to update to either $x_f$ or $x_f+1$. Because the rates are the same, the leader has time to randomly dance back and forth before the follower takes a step forward. The dynamics of the follower thus approximate the perfectly biased random walk, but the leader has additional dynamics. These additional dancing dynamics make the $\Gamma_f = \Gamma_\ell$ case less analytically tractable in the many-body setting below, but still result in clearly active dynamics as we show numerically.

As long as we keep $J \gg T_\ell, T_f$, the model retains a net current even as we relax the other limits. The diffusion remains biased except at $\epsilon=0$ or $T_f=T_\ell$, with the sign of the velocity equal to the sign of $(\epsilon \cdot \delta T)$, where $\delta T= T_\ell - T_f$ measures how far out of equilibrium the system is.

\textbf{The Brownian pump} is a more recognizably driven system: a thermal system relaxing according to a time-dependent Hamiltonian coupled to a cold bath at temperature $T$. 
The periodic time dependence drives the system, injecting energy and allowing for steady-state currents, while the cold bath ensures that the system autonomously relaxes toward the instantaneous ground state.
Prior works have studied transport in pump systems~\cite{Rahav2008, Asban2014} and shown that pumps are universal in the few-body setting~\cite{Raz2016}, but our construction here will be more generalizable to the many-body case.

\begin{figure}
	\centering
	\includegraphics[]{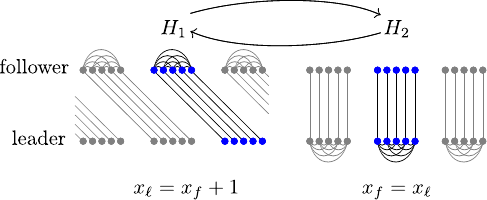}
	\caption{Arrangement and coupling of leader and follower spins in the Brownian pump for biased diffusion. Blue disks represent the positions of the particles in an instantaneous ground state.
    In $H_1$, the follower spins are all coupled to each other, and the leader spins are coupled individually such that $x_\ell=x_f+1$. In $H_2$,
    the leader spins are all coupled to each other, and the follower spins are coupled individually such that the ground state is $x_f = x_\ell$.}
	\label{fig:pump}
\end{figure}

The setup is illustrated in Fig.~\ref{fig:pump}, with two types of particles: leaders and followers. There are $n>2$ of each type for some small finite $n$ (with $n=5$ in the figure) indexed by $\alpha$. Let all particles start in position $x=0$. For the first half of the evolution, we turn on a Hamiltonian $H_1$ whose ground state has the leaders one step ahead of the followers. The cold bath makes the spins want to relax toward this ground state. We want to make sure that the leaders step forward, not that the followers step back. To make sure that the followers stay in place, we turn on an additional term in the Hamiltonian that makes it energetically favorable for all of the followers to be in the same position. This term penalizes any individual follower spin changing its state relative to the other followers, implementing a \emph{local energy barrier} for any individual follower spin to move. For the follower spins to move without violating this energy term, they would all have to move simultaneously. On the other hand, individual leader spins are free to update into their local ground state. This effectively makes the followers heavier, slowing their dynamics, so that the leader are more likely to step forward first. 

During the second half of the evolution, we turn on a Hamiltonian $H_2$ that instead makes the leaders heavy, by preferring all of the leaders to be in the same position. The Hamiltonian also prefers the leaders and followers to be in the same position, so that the followers step forward to match the leaders. In both cases, the apparent nonreciprocity, where one species of particle is more likely to move than the other, is made possibly by having $n>1$ particles of each type and coupling them together.

The Hamiltonian for this procedure  is periodic with period $2\tau$, $H(t+2\tau)= H(t)$, and given by 
\begin{equation}
	H(t) = \begin{cases}
		H_1, & 0 \le  t  < \tau \\
		H_2, & \tau\le  t  < 2\tau,
	\end{cases}
\end{equation}
with 
\begin{equation}
	H_1 = -J \sum_{\alpha < \beta}\delta(x_{f, \alpha}, x_{f, \beta}) - J \sum_\alpha \delta (x_{\ell,\alpha}, x_{f, \alpha}+1),
\end{equation}
and 
\begin{equation}
	H_2 = -J \sum_{\alpha < \beta}\delta(x_{\ell, \alpha}, x_{\ell, \beta}) - J \sum_\alpha  \delta (x_{\ell, \alpha}, x_{f, \alpha}),
\end{equation}
where $\alpha$ and $\beta$ run from 1 to $n$ and the $\delta$ functions are taken mod $L$. The Hamiltonian is illustrated in Fig.~\ref{fig:pump}: In $H_1$, the first term (represented by curved lines) holds the follower spins fixed while the second term (diagonal lines) moves the leaders one step ahead of the followers. In $H_2$, the first term (curved lines) holds the leader spins fixed while the second term (vertical lines) updates the followers to match the leaders.

The system approaches perfect biased diffusion in the limit
\begin{equation}
	1 \ll \tau \ll \exp (\beta J) ,
\end{equation}
where time is measured in units such that any given particle updates once on average over a time interval of unit length. 
Note that this implies $T \ll J$.
We can explain the two sides of the limit separately. First, we have $1 \ll \tau$ to make sure that we hold the Hamiltonian long enough for the leader to update or for the follower to match. Second, we have $\tau \ll  \exp (\beta J)$ so that we don't hold the Hamiltonian for so long that the particles become thermally excited. The result is that all particles update to $x(t+2\tau)=x(t)+1$ mod $L$: maximally biased diffusion.

If we relax the $1 \ll \tau$ limit while keeping $\tau \ll e^{\beta J}$, we will not get perfectly biased diffusion. Instead, there will be some probability that not all spins update each time step, so that the diffusion remains biased, with $p< 1$ but $q=0$ still.

\section{Summary of results} \label{sec:summary}

The remainder of this work explores generalizations of the above examples to general many-body spin systems. Our paper contains three main results. The first two are rigorous theorems concerning the simulation of probabilistic cellular automata either in continuous time with asynchronous updates or in discrete time with synchronous updates.\footnote{Note that this includes deterministic CAs as a special case.}
The third is a detailed study of an example that we find particularly illustrative. We summarize these results below. 

\textbf{1. Many-body Brownian ratchets can simulate any continuous-time PCA.} 

Consider continuous-time PCA dynamics $\mca_\async$, where individual spins update in continuous time according to a local rule. 
These are also called asynchronous PCA dynamics because there is no global clock to synchronize the individual updates.
For any PCA, we construct a ratchet system $\mcr$ whose dynamics closely approximates the dynamics of $\mca_\async$. The ratchet consists of
one copy of the spins (the leaders) evolving at temperature $T_\ell$ and clock rate $\Gamma_\ell$, and a second copy of the spins (the followers) evolving at temperature $T_f$ and clock rate $\Gamma_f$, coupled by a fixed Hamiltonian H.
We then prove that, when the system is updated according to Glauber dynamics,\footnote{The choice of Glauber dynamics is somewhat arbitrary, and we believe that other choices of thermal dynamics would lead to qualitatively similar constructions but would change some details.} the leader spins
have flip probabilities that are the same as those of $\mca_\async$, up to a controllable error: 

\begin{result}[Brownian Ratchets simulate asynchronous PCA dynamics]
    For every asynchronous PCA dynamics $\mca_\async$, there exists a Brownian ratchet $\mathcal R$ with Hamiltonian $H$ that simulates $\mca_\async$ up to error $\delta$, i.e. the transition rates of $\mca_\async$ and $\mathcal R$ differ by at most $\delta$, where $\delta$ can be made arbitrarily small by tuning system parameters such as energies, temperatures, and clock rates. 
\end{result}
For a mathematically precise statement, see Thm.~\ref{thm:ratchet}.

\textbf{2. Many-body Brownian pumps can simulate any discrete-time PCA.}
Now consider discrete-time PCA dynamics $\mca_\sync$, where all spins update in parallel according to a local rule. These are also called synchronous PCA rules because all spins update simultaneously.
For any PCA, we similarly construct a pump $\mathcal{P}$ with a time-periodic Hamiltonian $H(t) = H(t + k\tau)$, where $k=2$ for deterministic CAs and $k=3$ for PCAs,
such that thermal dynamics under $H(t)$ at an appropriate inverse temperature $\beta$ mimics the dynamics of $\mca_\sync$ up to a controllable error. 
Like the few-body pump in the previous section, we start with $2n$ copies of the system, with $n$ odd to simplify the proof, divided up into leaders and followers. 
The transition probability of $\mcp$ is the probability that the leader spins at a given site flip the value of their majority vote between time $t$ and time $t + k\tau$. We then show: 
\begin{result}[Brownian pumps simulate synchronous PCA dynamics]
For any synchronous PCA dynamics $\mca_\sync$, there exists a Brownian pump $\mathcal P$ that simulates $\mca_\sync$ up to error $\delta$, i.e. the transition probabilities of $\mca_\sync$ and $\mathcal P$ differ by at most $\delta$, where $\delta$ can be made arbitrarily small by tuning system parameters such as energies, temperature, and drive frequency.
\end{result}
For a mathematically precise statement, see Thm.~\ref{thm:pump}.

\begin{figure}
    \centering
    \includegraphics[]{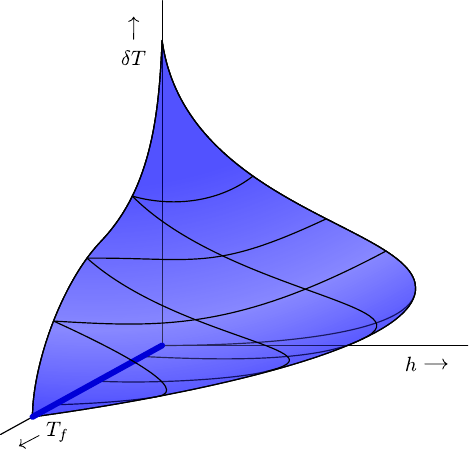}
    \caption{Sketch of the phase diagram of the ferromagnetic Toom ratchet as a function of the strength $h$ of a symmetry-breaking field, the temperature difference $\delta T$ between layers, and the follower temperature $T_f$. An equilibrium Ising model is not stable to any symmetry-breaking field, leading to a paramagnetic phase at all nonzero values of $h$ when $\delta T = 0$ as shown by the blue line. However, when we raise $\delta T$, the ferromagnetic phase (shaded in blue) extends to nonzero $h$. The upper limit to the ferromagnetic phase in the $T_f, \delta T$ plane are set by the energy scales in the Hamiltonian.}
    \label{fig:active_phase}
\end{figure}

\textbf{3. A simple two-temperature nearest-neighbor Ising model can be stable to a symmetry-breaking field.} While the constructions behind the previous two general results are not particularly complicated, they do involve multi-spin interactions between spins and (potentially all of) their neighbors. For specific examples, simpler Hamiltonians are possible. To illustrate this for a ratchet-based simulation of Toom's rule, we build a nearest-neighbor two-layer Ising model whose ferromagnetic phase (capable of supporting a memory) is stable to a symmetry-breaking field, as long as the different layers are held at different temperatures. The phase diagram sketched in Fig.~\ref{fig:active_phase}, with the temperature difference $\delta T = T_\ell-T_f$ on the vertical axis, the field $h$ on the horizontal axis, and the follower temperature $T_f$ on the third axis.

For pedagogical reasons, we present the simple ratchet example based on Toom's rule first, in Sec.~\ref{sec:ratchet}. We then present the generic ratchet construction in Sec.~\ref{sec:generic}, and the generic pump construction in Sec.~\ref{sec:pumps}. 

\section{Brownian Toom Ratchet} \label{sec:ratchet}

In this section, we realize Toom-like dynamics in a simple nonequilibrium Ising model. There are ferromagnetic couplings of two strengths, coupling some spins to a hot bath and some spins to a cold bath.  Previous work in this field has studied Ising models with temperatures that vary in time~\cite{Garrido1987, Tamayo1994, Beyen2024} or in space~\cite{lopez2006probabilistic, Khodabandehlou2024}. 
The closest work is Ref.~\cite{Blote1990}, which also studies an Ising model with two sublattices at different temperatures. Additionally, Ref.~\cite{Nakajima2008} studies a spin-lattice-gas model with different degrees of freedom at different temperatures.
However, none of this previous work explores the stability of the ferromagnetic phase to a symmetry-breaking field.  
In this language our result shows that the ferromagnetic phase of nonequilibrium Ising models can be robust even to symmetry-breaking fields.

\subsection{Review of Toom's rule}

Toom's rule~\cite{Toom1980} is a 2D cellular automaton with a memory time that scales exponentially in the linear system size $L$. 
Like the Ising model, it has two uniform steady states. 
Unlike the Ising model, each spin only tries to align with its north and east neighbor instead of trying to align with all four of its nearest neighbors. This means that domain walls move ballistically towards the southwest, unlike the curvature-driven dynamics of domain walls in the Ising model. In turn, this makes finite-size domains \emph{erode}, or disappear, in a time linear in the size of the domain instead of quadratic. The asymmetric nature of the update rule means that the interactions are inherently \emph{nonreciprocal}: each spin is influenced by its north and east neighbors, but doesn't influence them back. Since any local Hamiltonian will induce reciprocal interactions in equilibrium, nonreciprocity is a clear signature of active dynamics~\cite{FruchartNonreciprocity}. 

Due to the linear erosion, the memory time of Toom's rule remains exponential in $L$
even under symmetry-breaking noise that is more likely to flip to a particular value of the spins (which we will refer to as ``biased'' noise in the following). 
The noise bias is the analogue of a symmetry-breaking field in the Ising model because it prefers one of the steady states over the other.
While Toom's rule itself is a deterministic CA, the memory phase is also robust to probabilistic noise, when the dynamics are either synchronous~\cite{Toom1980} or asynchronous~\cite{gray1999toom, swart2026peierls}. 

Since our ratchet runs in continuous time, it will directly implement asynchronous dynamics. Furthermore, due to the thermal fluctuations of the two baths, the dynamics will be perturbed by probabilistic noise. These are the settings where Toom's rule is proven to be stable, 
so our goal in this section is to build a nonequilibrium Ising model that is stable to a symmetry-breaking field by simulating Toom's rule.

\subsection{Ratchet Hamiltonian}

\begin{figure}
	\centering
	\includegraphics[]{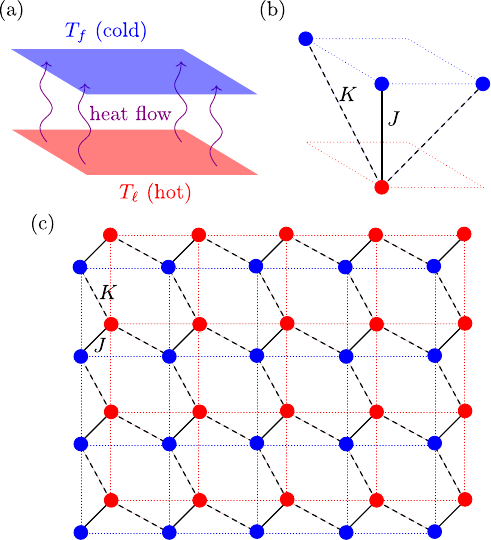}
	\caption{ 
    Three different representations of the Toom ratchet. Leader spins are red (for hot) and follower spins are blue (for cold). (a) Two square lattices stacked on top of each other, with heat flowing between the layers. (b) The couplings of a single leader spin, with solid lines for strength $J$ and dashed lines for strength $K$. Dotted lines are only a guide to the eye. (c) An overhead view emphasizes that the two layers can be reinterpreted as two sublattices at different temperatures.
    }
	\label{fig:honeycomb}
\end{figure}

Taking inspiration from the single-body ratchet, we consider a doubled square lattice of spins, with leader spins $\sigma_\ell$ and follower spins $\sigma_f$. The Hamiltonian is
\begin{align}
	H_\mathrm{Brownian} &= -J \sum_{i} \sigma_{\ell, i} \sigma_{f, i} \nonumber\\ 
	&\quad - \frac{K}{2} \sum_{i} \big( \sigma_{\ell, i} \sigma_{f, i+\hat{x}} + \sigma_{\ell, i} \sigma_{f, i+\hat{y}} \big) \nonumber\\
    &\quad - h \sum_i (\sigma_{\ell, i} + \sigma_{f, i}),
\end{align}
with $J = K+\epsilon$ and $T_\ell = T_f + \delta T$, where we set $h=0$ until Sec.~\ref{sub:field}. The leader spins are connected to a bath at temperature $T_\ell$ and update at clock rate $\Gamma_\ell$, and the follower spins are connected to a bath at temperature $T_f$ and update at clock rate $\Gamma_f$.  See Fig.~\ref{fig:honeycomb} for a graphical representation. In the remainder of this subsection we will analyze various limits to see how the couplings in the middle line, between $\sigma_{\ell, i}$ and its north and east follower neighbors, implement Toom's rule. In Sec.~\ref{sub:nofield} and~\ref{sub:field} we numerically explore the phase diagram when all parameters are of the same magnitude.

The heat engine implementation simulates Toom's rule arbitrarily well in the limit
\begin{equation}
	T_f \ll \epsilon \ll T_\ell \ll J, \quad \Gamma_f \gg \Gamma_\ell,
\end{equation}
To analyze this limit, first consider Glauber updates of the follower spins. Since $T_f$ is much smaller than the difference between $J$ and $K$, $\sigma_{f, i}$ always wants to match $\sigma_{\ell, i}$. Furthermore, since $\Gamma_f \gg \Gamma_\ell$ the system quickly reaches a state where $\sigma_{f,i}=\sigma_{\ell, i}$ for all $i$. When $\sigma_{\ell, i}$ updates, its temperature is high enough to not see the energy scale $\epsilon$. It feels the attraction of three follower spins: $\sigma_{f,i}$, $\sigma_{f,i+\hat{x}}$, and $\sigma_{f,i+\hat{y}}$. If $\sigma_{f,i+\hat{x}} = \sigma_{f,i+\hat{y}} \neq \sigma_{f,i}$---the only situation in which the Toom update would flip the spin at site $i$---then $\sigma_{\ell, i}$ will flip with probability close to $0.5$. In all other situations, $\sigma_{\ell, i}$ flips with a probability that scales as $e^{-J / T_\ell}$. 
In other words, the leader spins undergo Toom dynamics after marginalizing over the follower spins. In situations where the Toom update (applied only to the leader layer) would leave the spin as is, it does so with high probability. In situations where the Toom update (applied only to the leader layer) would flip the spin, it does so with probability $0.5$, and the follower updates to match nearly immediately.

Consider keeping the $T_f \ll \epsilon \ll T_\ell \ll J$ limit but set $\Gamma_f = \Gamma_\ell$. In situations where $\sigma_{\ell, i}$ is supposed to flip, it still does so with probability $\frac12$. However, because $\Gamma=1$, it now has time to dance back and forth, proposing and undoing the Toom update repeatedly, before the follower updates and accepts the update. In this limit, the dynamics remain highly nonequilibrium, but are not perfect Toom dynamics because of the many steps back and forth.

When $T_f=T_\ell$, the system is an equilibrium Ising model, and $\epsilon$ just parametrizes the difference in couplings on different bond types. On the other hand, if we keep the limits 
\begin{align}
	T_f \ll J,\, T_\ell,\, \epsilon,
\end{align}
without enforcing $\epsilon \ll T_\ell \ll J$, then the follower spins always try to match the leader spins but the leader spins do not follow perfect Toom updates. This is the limit in which the system looks most like an ordinary noisy Toom's rule CA. Thus, our model allows us to realize a version of the phase diagram in Fig. 3 of Ref.~\cite{rakovszky2024defining}, interpolating between Toom and Ising updates. 

Because Glauber updates are proposed at independently random times on each site, there is no obvious way to simulate synchronous (discrete-time) Toom dynamics. This should not be too surprising because there is no ``global clock'' to break the continuous time-translation symmetry.\footnote{A synchronous automaton can easily be simulated by an asynchronous one by coupling to a small additional noiseless automaton; see e.g. Ref.~\cite{berman1988investigations, LakeDecoder}. Relaxing the assumption that the additional automaton be noiseless appears to be difficult in general \cite{cook2008self}.}

The Hamiltonian with $h=0$ has two $\mathbb{Z}_2$ symmetries. The first is simply the Ising symmetry that flips all of the spins. When $T_f=T_\ell$ is smaller than the scale set by $J$ and $K$, this symmetry is spontaneously broken and leads to an equilibrium classical memory. The other symmetry consists of spatial inversion about the origin in each layer, combined with swapping $T_f$ and $T_\ell$ and relabeling the leader and follower spins.
This means that when $T_f$ is larger than $T_\ell$, the simulation runs in the opposite spatial direction. 

We can add additional intra-layer coupling to improve the memory, but choose not to in order to focus on the active dynamics. Conversely, we can also remove equilibrium ferromagnetic order completely so that the system never orders when $T_f=T_\ell$ and the memory only persists when the baths are at different temperatures 
(see Sec.~\ref{sub:glauber}).

\subsection{Clock rates and emergent nonreciprocity}

Toom's rule requires that each spin is influenced by its north and east neighbors, but does not influence them back. On the other hand, for any dynamics that obey detailed balance with respect to a free energy, any pair of spins must influence each other in the same way. Thus, the notion of nonreciprocal interactions is central to active dynamics~\cite{FruchartNonreciprocity, Nadolny2025}. 
The many-body  ratchet construction allows us to see the nonreciprocity emerge from a reciprocal Hamiltonian. The derivation of this \emph{emergent nonreciprocity} is particularly clear in the limit $\Gamma_f \gg \Gamma_\ell$, because we can integrate out the fast followers. We emphasize that $\Gamma_f \gg \Gamma_\ell$ is just convenient for analysis, and the amount of nonreciprocity is controlled by $\delta T$.

In the limit $\Gamma_f \gg \Gamma_\ell$, we can integrate out the fast follower spins. Write the Hamiltonian as $H(\sigma_\ell, \sigma_f)$ where $\sigma_\ell$ denotes the entire leader configuration and $\sigma_f$ denotes the entire follower configuration.  If $\sigma_\ell^i$ denotes the leader configuration obtained from $\sigma_\ell$ by flipping $\sigma_{\ell, i}$, then let 
\begin{equation}
\Delta_iH(\sigma_\ell, \sigma_f) = H(\sigma_\ell^i, \sigma_f) - H(\sigma_\ell, \sigma_f)
\end{equation}
denote the energy difference due to flipping $\sigma_\ell^i$. For a fixed leader configuration $\sigma_\ell$, the followers rapidly relax to the conditional Boltzmann distribution
\begin{align}
    P_{\beta_f}(\sigma_f | \sigma_\ell) &= \frac{e^{-\beta_f H(\sigma_\ell, \sigma_f)}}{Z_f(\sigma_\ell)}, \\
    Z_f(\sigma_\ell) &= \sum_{\sigma_f} e^{-\beta_f H(\sigma_\ell, \sigma_f)}.
\end{align}
The effective flip rate for $\sigma_{\ell, i}$ is then
\begin{align}
W_i(\sigma_\ell) &= \frac{1}{Z_f(\sigma_\ell)} \sum_{\sigma_f} e^{-\beta_f H(\sigma_\ell, \sigma_f)}  w_i(\sigma_\ell | \sigma_f),
\end{align}
where 
\begin{equation}
w_i(\sigma_\ell | \sigma_f) = \frac{1}{1+e^{\beta_\ell \Delta_i H(\sigma_\ell, \sigma_f)}}
\end{equation}
is the Glauber flip rate for $\sigma_{\ell, i}$.

To see the violation of detailed balance, define $\delta \beta = \beta_f - \beta_\ell$. Using the Glauber identity 
\begin{align}
e^{-\beta_\ell H(\sigma_\ell, \sigma_f)} w_i(\sigma_\ell | \sigma_f) = e^{-\beta_\ell H(\sigma_\ell^i, \sigma_f)} w_i(\sigma_\ell^i | \sigma_f),
\end{align}
we can rewrite the forward rate as
\begin{equation}
    W_i(\sigma_\ell) = \frac{1}{Z_f(\sigma_\ell)} \sum_f e^{-\delta\beta H(\sigma_\ell, \sigma_f)} e^{-\beta_\ell H(\sigma_\ell^i, \sigma_f)} w_i(\sigma_\ell^i | \sigma_f),
\end{equation}
and the reverse rate as
\begin{equation}
    W_i(\sigma_\ell^i) = \frac{1}{Z_f(\sigma_\ell^i)} \sum_f e^{-\delta\beta H(\sigma_\ell^i, \sigma_f)} e^{-\beta_\ell H(\sigma_\ell^i, \sigma_f)} w_i(\sigma_\ell^i | \sigma_f).
\end{equation}
Therefore, their ratio is
\begin{equation}
\frac{W_i(\sigma_\ell)}{W_i(\sigma_\ell^i)} = \frac{Z_f(\sigma_\ell^i)}{Z_f(\sigma_\ell)} \frac{\sum_f e^{-\delta\beta H(\sigma_\ell, \sigma_f)} e^{-\beta_\ell H(\sigma_\ell^i, \sigma_f)} w_i(\sigma_\ell^i | \sigma_f)}{\sum_f e^{-\delta\beta H(\sigma_\ell^i, \sigma_f)} e^{-\beta_\ell H(\sigma_\ell^i, \sigma_f)} w_i(\sigma_\ell^i | \sigma_f)}
\end{equation}
which we can compare to equilibrium expectations.

When the two temperatures are equal, $\delta \beta = 0$, the second factor is exactly one. This allows us to write
\begin{equation}
\frac{W_i(\sigma_\ell)}{W_i(\sigma_\ell^i)} = \frac{Z_f(\sigma_\ell^i)}{Z_f(\sigma_\ell)} = e^{-\beta\left[F_{\rm eff}(\sigma^i_\ell)-F_{\rm eff}(\sigma_\ell) \right]},
\end{equation}
where
\begin{equation}
     F_{\rm eff}(\sigma_\ell) = -\frac{1}{\beta} \log Z_f(\sigma_\ell)
\end{equation}
is the effective free energy. Thus, whenever $T_\ell = T_f$, the leader dynamics is generated by reciprocal effective interactions.

When $T_f \ne T_\ell$, however, the second factor is generally nontrivial,
and depends on the particular move $\sigma\to\sigma^i$, not only on the initial and final partition functions.
Generically it cannot be absorbed into a scalar effective free energy
for the leaders alone because the followers are distributed according to $\beta_f$ while leader flips are accepted according to $\beta_\ell$.  This mismatch is the source of the nonequilibrium, nonreciprocal effective leader dynamics.

A similar analysis applies if we instead let the leaders update much faster than the followers. In fact, in our numerics we set the update rates to be equal and still find that the ferromagnetic phase is stable to a symmetry-breaking field. The fact that the ratchet evades equilibrium constraints even when the updates are equal means that in this case, the effective interactions must still be nonreciprocal.

\subsection{Phase diagram with Ising symmetry} \label{sub:nofield}

The next question is when the simulation is ``good enough'' to function as a classical memory in the absence of a symmetry-breaking field ($h=0$). In other words, to what extent to the properties of Toom's rule persist when the parameters $J$, $K$, $T_\ell$, $T_f$, $\Gamma_\ell$, and $\Gamma_f$ are all of the same order of magnitude? To answer this question, we turn to single-spin-flip Glauber dynamics with $J=1$, $K=0.8$, and $\Gamma_\ell = \Gamma_f = 1$.

We expect to find two phases, one where the model is a good classical memory (analogous to a ferromagnet) and one where it is not (analogous to a paramagnet). In order to distinguish these, we start the system with all spins down, evolve for a time long enough to reach the local steady state of the dynamics, and then measure the magnetization. In the good memory phase the system remains in the mostly down state while in the bad memory phase the system approaches a state that has magnetization 0. In this experiment, we do not have to worry about metastability because the all-up and all-down states should have the same lifetime due to Ising symmetry. 

\begin{figure}
	\centering
	\includegraphics[]{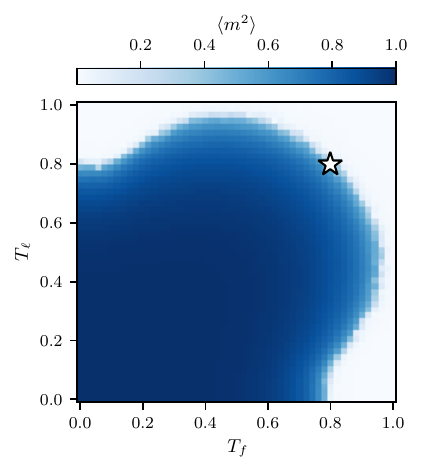}
	\caption{Steady-state phase diagram of the Toom ratchet at $J=1$, $K=0.8$, with no field. The value displayed is the expectation value of magnetization squared, so that a value of 0 is the paramagnetic phase and 1 is the ferromagnetic phase.  The phase transition is equilibrium Ising on the diagonal $T_f=T_\ell$ and is most Toom-like on the axes. The star marks the exact location of the equilibrium transition.}
	\label{fig:equilibrium}
\end{figure}

In particular, we work on an $L=100$ system. We thermalize the system for 10,000 time steps and then take 1,000 samples each separated by 100 time steps. We use $\langle m^2 \rangle$ as our order parameter, which should vanish in the paramagnetic phase and be nonzero in the ferromagnet phase. 
The results are shown in Fig.~\ref{fig:equilibrium}. We see a ferromagnetic phases when both temperatures are small and a paramagnetic phase when either temperature is high enough.

As a check, we can calculate the critical temperature exactly when $T_f = T_\ell$, because this is an ordinary equilibrium Ising model. First, use the condition
\begin{align}
	1 &= \phantom{+} \tanh(\beta_c J_1) \tanh(\beta_c J_2) \nonumber\\
	&\quad + \tanh(\beta_c J_2) \tanh(\beta_c J_3)  \nonumber\\
	&\quad + \tanh(\beta_c J_3) \tanh(\beta_c J_1) 
\end{align}
for the critical temperature of the anistropic honeycomb Ising model~\cite{Baxter}. For our purposes, $J_1=J$ and $J_2 = J_3 = K/2$. We can then numerically solve for the critical temperature when $J=1$ and $K=.8$, finding
\begin{equation}
	T_c \approx .799,
\end{equation}
which we mark with a star in Fig.~\ref{fig:equilibrium}. This point does, in fact, fall close to the apparent phase boundary.

\subsection{Phase diagram with a symmetry-breaking field} \label{sub:field}

We now want to test the memory in the presence of the perturbation
\begin{equation}
	\delta H = - h \sum_i \big(\sigma_{\ell, i} + \sigma_{f, i} \big),
\end{equation}
a symmetry-breaking field. Any equilibrium Ising model can not be stable with respect to such a field, so stability serves to show that our model does not approach equilibrium \emph{and} that it avoids equilibrium in a ``useful'' way.

\begin{figure}
	\centering
	\includegraphics[]{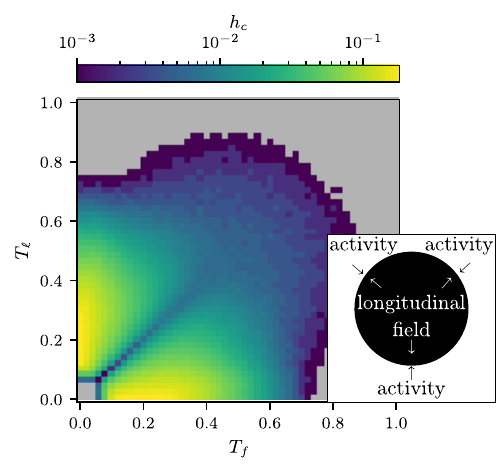}
	\caption{Critical field from erosion experiment, on log scale. The region of stability appears to be slightly smaller than in Fig.~\ref{fig:equilibrium}, but this is likely due to a combination of finite-size effects and a very small critical field. Furthermore, although the critical field appears to be nonzero along the diagonal, we know this cannot be the case and must be due to finite-size effects. [Inset] A symmetry-breaking field applies pressure to grow the island while the active dynamics try to shrink it.
	}
	\label{fig:erosion_log}
\end{figure}

Distinguishing the two phases is now more difficult because of the metastability of the unfavored state. In the memory phase, the unfavored state has a lifetime that is exponentially long in $L$. Outside of the memory phase, the lifetime of the unfavored state becomes finite due to the formation of bubbles of the preferred state that grow to cover the whole system. However, near the critical field strength $h_c$, this lifetime can be exponentially large in $1/|h-h_c|$, so that even though it is finite it is still numerically large.\footnote{Compare this to the equilibrium Ising model, where $h_c=0$ and the metastable state still has a parametrically large lifetime, $\tau\sim \exp (1/|h|)$~\cite{langer1969theory}.}

In order to approach this problem we analyze the erosion property of Toom's rule: Starting in a disfavored state, a finite island of the preferred state should be removed in finite time. Then, for fixed $J$, $K$, $T_\ell$, and $T_f$, we define the \emph{apparent critical field} as the largest field for which the active dynamics can still win and shrink the circle (see inset in Fig.~\ref{fig:erosion_log}). Note that this experiment is susceptible to finite-size effects: for a fixed bubble size there is a critical field strength below which the energetic penalty from the boundary will win, even without activity. Thus, our apparent critical field can differ from the true critical field by an amount of order $1/L$. In comparison, for an equilibrium Ising model the erosion experiment will find an apparent critical field of order $1/L$ because the true critical field is 0.

We plot the critical field against $T_\ell$ and $T_f$ in Fig.~\ref{fig:erosion_log}.
Note that the apparent critical field does not exactly vanish for all $T_f=T_\ell$ because of the finite-size effects. Similarly, finite-size effects cause the small grey square at small $T$ because at very small temperatures the bubble neither grows nor shrinks. 

Nevertheless, the pattern is clear: The system only has a nonzero critical field within the ordered phase (compare to Fig.~\ref{fig:equilibrium}. Furthermore, within that phase, the critical field is largest when $T_f$ is very small and $T_\ell$ is near $0.1$, which is comparable to $\epsilon = 0.2$. The phase diagram is symmetric upon exchanging $T_f$ and $T_\ell$, as expected from the symmetries of the Hamiltonian.

\begin{figure}
	\centering
	\includegraphics[]{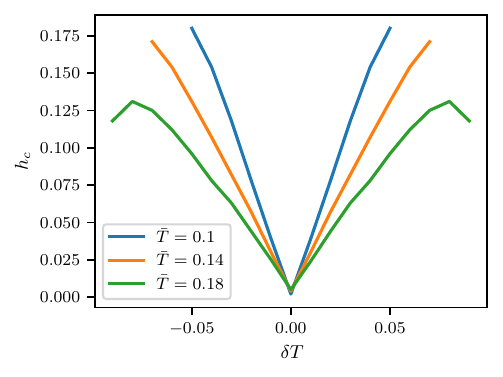}
	\caption{Apparent critical field vs $\delta T \equiv T_\ell-T_f$ for various values of $\bar{T} \equiv \frac12 (T_\ell+T_f)$. For all values of $\bar{T}$ the field is approximately linear, for small $\delta T$, with different proportionality constants. }
	\label{fig:linear_hc}
\end{figure}

We can also ask how the critical field depends on $\delta T = T_\ell-T_f$ when the temperature difference is very small and the sum of temperatures is held constant. Linear response suggests that the critical field should be linear in $\delta T$, and Fig.~\ref{fig:linear_hc} shows that this is correct. 

\section{Many-body ratchets} \label{sec:generic}

In this section we give a generic ratchet construction for the
asynchronous dynamics generated by a completely general PCA \(A\). We use ``asynchronous PCA dynamics'' to mean the continuous-time local Markov process $\mathcal A_\async$ obtained by applying a local PCA rule to each site at independent Poisson-distributed update times. At each site $i$, the probability $p_i^\mca$ for a spin to flip depends only on the spins in some neighborhood $\mathcal{N}(i)$. The ratchet again consists of one copy of the system (the leaders) evolving at temperature $T_\ell$ and clock rate $\Gamma_\ell$, and a second copy of the system (the followers) evolving at temperature $T_f$ and clock rate $\Gamma_f$, coupled by a fixed Hamiltonian. When the ratchet operates at finite temperatures and energy parameters, the simulation will be noisy. However, in the limit $T_f \ll \epsilon \ll T_\ell \ll J$ and $\Gamma_f \gg \Gamma_\ell$ we will approach error-free (but still asynchronous) operation. 

\subsection{Definitions}

Consider a collection of spins whose state is given by $\sigma \in \{\pm 1\}^N$. While some PCAs like Toom's rule have deterministic update rules, it is also useful to consider generic PCAs where, even before noise is added in, some flip probabilities are between 0 and 1. As an example, the squeezing codes from Ref.~\cite{lake2025squeezing} are probabilistic and realize critical exponents that do not correspond to any known equilibrium model, whereas Toom's rule has the same critical exponents as the Ising model.

\begin{definition}[Probabilistic cellular automaton]
A probabilistic cellular automaton $A$ is specified by the flip probabilities 
\begin{equation}
p_i^A(\sigma) \in [0,1],
\end{equation}
where $p_i^A(\sigma)$ depends only on the spins in some neighborhood $\mathcal{N}(i)$ with size $|\mathcal{N}|=r$. The PCA is a deterministic CA if all $p_i^A(\sigma)\in \{0,1\}$.
\label{def:CA}
\end{definition}

Given the flip probabilities, the dynamics corresponding to the PCA can be either asynchronous and in continuous time, or synchronous and in discrete time. We treat the asynchronous case here and delay the synchronous case to Sec.~\ref{sec:pumps}.
\begin{definition}[Asynchronous PCA dynamics]
Given a PCA $A$, let $\mca_\async$ denote the local continuous-time Markov process defined by giving each spin an independent Poisson clock. When the clock at site $i$ rings, $\sigma_i$ flips with probability $p_i^A(\sigma)$. Equivalently, the spin-flip rate is
\begin{equation}
    W_i^{\mca_\async}(\sigma) = \Gamma_i p_i^A(\sigma),
\end{equation}
where $\Gamma_i$ is the local clock rate.
\end{definition}

For our ratchet simulations, we will need to consider also a {\it lazy} version of $A$, denoted $\widetilde A$: 
\begin{definition}[Lazy PCA]
A lazy PCA $\widetilde A$ is defined from $A$ by having flip probabilities that are smaller by a constant factor of $1/2$:
\begin{equation} 
	p^{\widetilde A}_i(\s) = \frac12 p^A_i(\s).
\end{equation} 
We denote the corresponding asynchronous dynamics by $\widetilde \mca$.
\end{definition}
Clearly, simulating $\widetilde \mca_\async$ is just as good as simulating $\mca_\async$ as far as the physics is concerned (since $\mca_\async$ is continuous-time). 

We also define $\delta$-closeness as a way to compare two continuous-time Markov processes.
\begin{definition}[$\delta$-closeness]
    Two local continuous-time Markov processes $\mathcal M$ and $\mathcal N$ are $\delta$-close if 
    \begin{equation}
        |W_i^\mathcal{M}(\sigma) - W_i^{\mathcal N}(\sigma)| \le \delta
    \end{equation}
    for all locations $i$ and configurations $\sigma$,
    where $W_i^{\mathcal M}(\sigma)$ are the rates corresponding to ${\mathcal M}$ and $W_i^\mathcal{N}(\sigma)$ are the rates corresponding to $\mathcal{N}$.
    \label{def:sim}
\end{definition} 

\subsection{Construction} \label{sub:ratchet}

For simplicity we will first construct the Brownian ratchet for a deterministic CA $A$, so that $p^{ A}_i(\s_f) \in \{ 0, 1 \}$.  The extension to PCA dynamics will be given in Sec.~\ref{sub:pcas}.
Taking inspiration from the previous constructions, we consider a doubled set of spins, with the leader spins $\sigma_\ell$ and the follower spins $\sigma_f$. 
The leader spins should update by applying the CA rule to the current state of the follower spins, and the follower spins should update to match the leaders. To do this, 
we create a Hamiltonian whose ground state is given by configurations where all leader-follower pairs match. Furthermore, we want to only lightly penalize non-matching configurations if an update is supposed to happen on site $i$ according to $A$, and strongly penalize them otherwise.
Such a Hamiltonian is 
\begin{align}
	H_A &= \sum_i J_i(\sigma_f) \Pi_i, \nonumber \\
	J_i(\sigma_f) &= \begin{cases}
		\epsilon, & p^{ A}_i(\s_f) = 1, \\
		J, &p^{ A}_i(\s_f) = 0,
	\end{cases} \label{eqn:levels}
\end{align}
where 
\begin{equation}
	\Pi_i = \frac{1-\sigma_{\ell, i}\sigma_{f, i}}{2}
\end{equation}
is a projector onto nonmatching configurations. 

\begin{figure}
    \centering
    \includegraphics[]{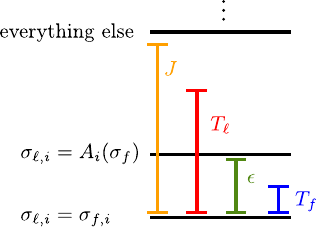}
    \caption{Energy levels in the many-body ratchet \eqref{eqn:levels}, showing that the leader spin is easily able to evolve forward in time via the CA, but the follower spin always just wants to match.
    }
    \label{fig:levels}
\end{figure}

The operation is particularly transparent in the  limit 
\begin{equation}
	T_f \ll \epsilon \ll T_\ell \ll J, \quad \Gamma_f \gg \Gamma_\ell
\end{equation}
illustrated in Fig.~\ref{fig:levels}.
To analyze this limit, first consider Glauber updates of the follower spins. Since $T_f\ll\epsilon$, $\sigma_{f, i}$ always wants to reach the true ground state by matching $\sigma_{\ell, i}$. On the other hand, when the leader $\sigma_{\ell, i}$ updates, its temperature is much less than $J$, so it never updates if $p_i^\mca(\sigma_f)=0$. On the other hand, 
since the leader temperature is much higher than $\epsilon$, when $p^{{A}}_i(\sigma_f) = 1$, then $\sigma_{\ell, i}$ flips with probability $\frac12$, hence simulating the update of the lazy CA. In turn, the leader flip proposes the update to the follower $\sigma_{f, i}$, and $\sigma_{f, i}$ immediately accepts the update because $\Gamma_f \gg \Gamma_\ell$. 

It is also instructive to rewrite the Hamiltonian in a different form. First, let us briefly abuse notation to write $A_i(\sigma) = \pm1$ to denote the value that $\sigma_i$ is supposed to update to. Then, up to constants, the Hamiltonian is 
\begin{align}
	H_A &= - J \sum_i \sigma_{\ell, i} \sigma_{f, i} \nonumber\\
	&\phantom{=} - K \sum_i \sigma_{\ell, i} A_i(\sigma_f) \nonumber \\
	&\phantom{=} + K \sum_i \sigma_{f, i} A_i(\sigma_f)  \label{eqn:antiferro}
\end{align}
with $K = J-\epsilon$. The first term makes $\sigma_{f, i}$ want to match $\sigma_{\ell, i}$ and the second term makes $\sigma_{\ell, i}$ want to update to $A_i(\sigma_f)$, the time-evolved value indicated by nearby $\sigma_f$. However, the second term can be dangerous because it makes nearby $\sigma_f$ want to be consistent with the current value of $\sigma_{\ell, i}$. The third term cancels this backreaction, as long as all of the nearby sites are matching.

Finally, let us define the Brownian ratchet as a continuous-time Markov process so that we can compare it to the asynchronous CA dynamics on equal footing.
\begin{definition}[Brownian ratchet]
For any PCA $A$, the Brownian ratchet dynamics $\mcr_A(J, \epsilon, T_\ell, T_f, \Gamma_\ell, \Gamma_f)$ is a local continuous-time Markov process defined by the Glauber dynamics induced on the leader spins by $H_A(J, \epsilon)$ (Eq.~\ref{eqn:levels} for deterministic CAs, Eq.~\ref{eqn:plevels} for generic PCAs), with leader spins updating at rate $\Gamma_\ell$ and temperature $T_\ell$, marginalizing over the follower spins updating at rate $\Gamma_f$ and temperature $T_f$. 
\end{definition}
We will treat the leader spins as the ones that are doing the simulation and treat the followers as an auxiliary system. 

\subsection{Brownian ratchet simulation} \label{sub:rates}

Having constructed the Brownian ratchet, we are prepared to state the simulation theorem.
\begin{theorem} [Brownian ratchets simulate continuous-time PCA dynamics]
    Let $A$ be a PCA with neighborhood size $\size$, and $\widetilde{\mca}_\async$ its lazy asynchronous dynamics. For every $\delta>0$, there exists a Brownian ratchet $\mcr_A(J, \epsilon, T_\ell, T_f, \Gamma_\ell, \Gamma_f)$ that is $\delta$-close to $\widetilde{\mca}_\async$. The simulation closeness is set by
    \begin{align}
    \delta \le \bigg( \frac{e^{\beta_\ell \epsilon} - 1}{2(e^{\beta_\ell \epsilon}+1)} + \frac{1}{1+e^{\beta_\ell J}} +& \nonumber \\ 
    \frac{\size \Gamma_\ell}{\size \Gamma_\ell + \Gamma_f} +& \frac{1}{1+e^{\beta_f \epsilon}} \bigg),
    \end{align}
    so that $\delta \rightarrow 0$ in the limit $T_f\ll\epsilon \ll  T_\ell \ll J$ and $\Gamma_f \gg \size \Gamma_\ell$. 
    \label{thm:ratchet}
\end{theorem}
We construct the proof via a union bound on the different possible error channels. We prove Thm.~\ref{thm:ratchet} for deterministic CA dynamics here, and generic PCA dynamics in the next subsection.

Consider a starting configuration where some sites are matching ($\sigma_{\ell, i} = \sigma_{f, i}$) and some are not. For the simulation to succeed, all of the leaders within the neighborhood of site $i$ need to wait until site $i$ is matching before they attempt an update. Since the followers update at rate $\Gamma_f$ and the leaders update at rate $\Gamma_\ell$ the probability that any of the nearby leaders attempt an update before $\sigma_{f, i}$ is
\begin{equation}
    p_\text{too soon} = \frac{\size \Gamma_\ell}{\size \Gamma_\ell + \Gamma_f}. \label{eqn:toosoon}
\end{equation}
Once $\sigma_{f, i}$ does update, it matches its leader with probability
\begin{equation}
    p_\text{match} = \frac{1}{1+e^{-\beta_f \epsilon}},
\end{equation}
because matching lowers the energy. Thus, the chance of finding a non-matching pair is $1-p_\text{match}$

Once all sites are matching, the energy is 0 everywhere. If a leader spin $\sigma_{\ell, i}$ is supposed to flip [$p^{\widetilde A}_i(\sigma_\ell)=.5$], it does so with probability 
\begin{align}
    p_\text{flip} = \frac{1}{1+e^{\beta_\ell \epsilon}},
\end{align}
so that
\begin{equation}
    p^{\widetilde \mca}_i(\sigma_\ell) - p_\text{flip} = \frac{e^{\beta_\ell \epsilon} - 1}{2(e^{\beta_\ell \epsilon}+1)}
\end{equation}
approaches $0$ as $\beta_\ell \epsilon \rightarrow 0$. If it is not supposed to flip, it does so with probability
\begin{equation}
    p_\text{error} = \frac{1}{1+e^{\beta_\ell J}},
\end{equation}
which approaches $p^{\widetilde \mca}_i(\sigma_\ell)=0$ as $\beta_\ell J \rightarrow \infty$.

Once $\sigma_{\ell, i}$ updates, $\sigma_{f, i}$ has to update before any of the nearby leaders attempt an update. As before, the probability that any of the nearby leaders attempt an update before $\sigma_{f, i}$ is $p_\text{too soon}$, and  once $\sigma_{f, i}$ does update, it matches its leader with probability $p_\text{match}$, so that the probability of finding  a non-matching pair is $1-p_\text{match}$.

Taking all of these error channels together, the many-body Brownian ratchet is a $\delta$-close simulation of the asynchronous lazy CA dynamics $\widetilde \mca$ with
\begin{align}
    \delta \le \bigg( \frac{e^{\beta_\ell \epsilon} - 1}{2(e^{\beta_\ell \epsilon}+1)} + \frac{1}{1+e^{\beta_\ell J}} +& \nonumber \\ 
    \frac{\size \Gamma_\ell}{\size \Gamma_\ell + \Gamma_f} +& \frac{1}{1+e^{\beta_f \epsilon}} \bigg),
\end{align}
by a union bound on the individual error rates,
so that $\delta \rightarrow 0$ in the limit $T_f \ll \epsilon \ll T_\ell \ll J$ and $\Gamma_f \gg r \Gamma_\ell$. This proves Thm.~\ref{thm:ratchet} for deterministic CAs.

\subsection{Intrinsically probabilistic CAs} \label{sub:pcas}

We now generalize the proof of Thm.~\ref{thm:ratchet} to generic PCAs. The Hamiltonian is 
\begin{align}
	H_A &= \sum_i J_i(\sigma_f) \Pi_i, \nonumber \\
	J_i(\sigma_f) &= \begin{cases}
		\epsilon, & p^{ A}_i(\s_f) = 1, \\
		J, &p^{ A}_i(\s_f) = 0, \\
        T_\ell \log \left( \frac{2}{p_i^{ A}(\sigma_f)} - 1 \right), &\text{otherwise},
	\end{cases} \label{eqn:plevels}
\end{align}
which generalizes the Hamiltonian from Sec.~\ref{sub:ratchet}.
We have already proven that the errors in the flip probability are controlled by Thm.~\ref{thm:ratchet} when $p^{ A}_i(\s_f) = 0, 1$, so all that remains is proving the bounds on the flip probability in the last case.

As we have already shown, the probability that a given leader tries to update before all of the followers in its neighborhood are in matching configurations is $p_\text{too soon}$~\eqref{eqn:toosoon}.
Then, for $p^{ A}_i(\s_f) \ne 0, 1$ the flip probabilty for $\sigma_{\ell, i}$ is
\begin{align}
    p_\text{flip} &= \frac{1}{1 + e^{\beta_\ell J_i(\sigma_f)}} \nonumber\\
    &= \frac{p_i^A(\s_f)}{2} = p_i^{\widetilde A}(\s_f),
\end{align}
with no limit needed. Combined with the results of the previous Subsection, this completes the proof of Thm.~\ref{thm:ratchet} for PCAs.

\subsection{Illustration: Toom's rule revisited} \label{sub:glauber}

\begin{figure}
	\centering
	\includegraphics[]{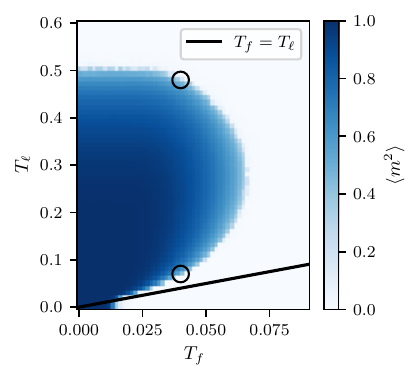}
	\caption{
    Steady-state phase diagram of the Toom ratchet derived from the general construction. The black line denotes equilibrium, $T_f = T_\ell$, which lies entirely within the disordered phase (at very low temperatures the ordered phase appears to push below the equilibrium line, but we believe this is just due to slow equilibration of the initial condition and a finite time window of the numerics). Black circles mark corresponding points on Fig.~\ref{fig:g_erosion}.
    }
	\label{fig:g_equilibrium}
\end{figure}

We can numerically explore the functioning of the many-body ratchet away from the regimes where we have analytical control. We will once again  use Toom's rule as our CA because it is stable to noise and to asynchronicity. Although we already constructed a many-body ratchet for Toom's rule in the previous Section, we find it useful here to apply the generic construction to Toom's rule to illustrate how the construction works, and find some different phenomenology.

As written,~\eqref{eqn:levels} applied to Toom's rule has an Ising spin-flip symmetry.
We expect to find two phases, one where the model is a good classical memory and one where it is not. When the Ising symmetry is intact, these two phases are just a ferromagnet and a paramagnet, so we can distinguish them by $\langle m^2 \rangle$, where $m$ is the average magnetization. 

In particular, we simulate single-spin-flip Glauber dynamics with $\Gamma_f = \Gamma_\ell =1$ for all spins. Working on an $L=100$ system, we thermalize the system for 10,000 time steps and then take 1,000 samples each separated by 100 time steps. The surprising result, as shown in Fig.~\ref{fig:g_equilibrium}, is that the ferromagnetically ordered phase only appears when $T_f<T_\ell$; the black line denotes equilibrium $T_f=T_\ell$. In other words, the antiferromagnetic couplings in~\eqref{eqn:antiferro} ensure that the model never orders in equilibrium, but only out of equilibrium. 

As in the previous section, we use an erosion experiment to test the memory in the presence of the perturbation
\begin{equation}
	\delta H = - h \sum_i \big(\sigma_{\ell, i} + \sigma_{f, i} \big),
\end{equation}
a symmetry-breaking field that favors $\sigma = \sgn(h)$.
Any equilibrium Ising model can not be stable with respect to such a field, so stability serves to show that our model does not approach equilibrium \emph{and} that it avoids equilibrium in a ``useful'' way.

\begin{figure}
	\centering
	\includegraphics[]{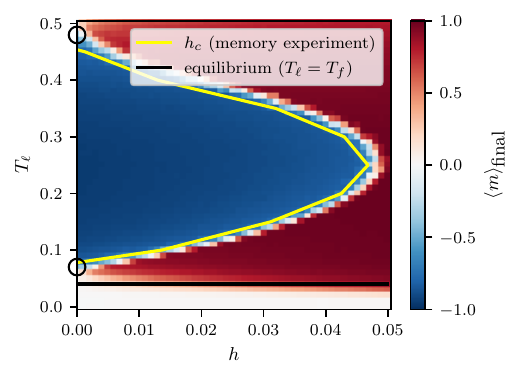}
	\caption{
    Late-time magnetization in the erosion experiment for the Toom ratchet derived from the general construction. Blue represents where the active dynamics are able to ``win'' and remove the initial domain. Red represents where the field wins, and white shows where the two effects are balanced or (at small $T_\ell$) where the temperature is too cold for any dynamics to happen within the timescale of the experiment.
	The yellow line is the phase diagram extracted from the memory experiment. Note the close agreement between the erosion and memory experiments, despite very different designs.
	Furthermore, note  the agreement with the steady-state phase diagram when they overlap, at $h=0$, $T_f=.04$, marked by black circles. 
    }
	\label{fig:g_erosion}
\end{figure}

We report the results of the erosion experiment at $L=100$, $J=1$, $K=.8$, $T_f=.04$ in Fig.~\ref{fig:g_erosion}. At each value of $T_\ell, h>0$ we initialize a +1 domain of size $L^2/2$ in a background of -1, so that the field prefers the domain, and plot the late time magnetization. If the symmetry-breaking field wins the late time magnetization is large and positive, and if the Toom dynamics wins the late time magnetization is large and negative. The white stripe, where neither wins, is the apparent critical field at each temperature. Unfortunately this system is similarly susceptible to finite-size and finite-time effects, as seen by the large band at small $T_\ell$ where the initial island neither grows nor shrinks, but the behavior at larger $T_\ell$ is clear.

As a complementary measure of the critical field, we can perform a full memory experiment and  evade the finite-size effects through scaling.
To do so, we start with all spins up and turn on a field that prefers spins down, and run the dynamics until a majority of the spins are down. The memory time defined this way should grow exponentially with $L$ in the good-memory phase. In the bad-memory phase, the memory time grows exponentially until a critical system size $L^*$, after which it decays to a plateau~\cite{Ray2024}. This behavior is also found in the Higgs phase of 3D gauge theory~\cite{Stahl2026} and in Haah's code~\cite{Bravyi2013}. 

\begin{figure}
	\centering
	\includegraphics[]{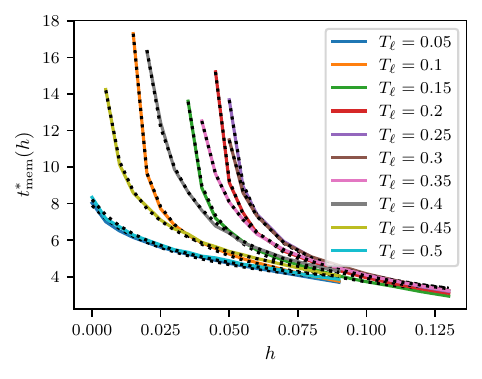}
	\caption{Divergence of $t_\mathrm{mem}^*$ with $h$, which we use to find $h_c$. The colored lines are data from the memory experiment and the dotted lines are the fits. We plot $h_c$ against $T_\ell$ in Fig.~\ref{fig:g_erosion}.}
	\label{fig:hc_fit}
\end{figure}

Working in the bad-memory phase, at fixed $J, K, T_\ell, T_f, h$, we numerically estimate $t_\textrm{mem}^*$, the best memory time achievable at those parameters. Then, by fitting $t_\textrm{mem}^*(h)$ to a power-law decay, we find $h^*(J, K, T_\ell, T_f)$, the critical value of $h$ where $t_\textrm{mem}^*$ diverges; see Fig.~\ref{fig:hc_fit}. This value defines the phase boundary. While the erosion experiment probes large system sizes and small times, the memory experiment probes smaller system sizes on much longer times.

To match the erosion experiment, we fix $J=1$, $K=.8$, and $T_f=.04$. We then plot the phase diagram on top of the erosion phase diagram in Fig.~\ref{fig:g_erosion}. 
Note the similarity between the two experiments, showing that the finite-size effects in the erosion experiment are not too strong and validating the erosion experiment. 

\section{Many-Body Brownian Pumps} \label{sec:pumps}

In this section we show how to replace the heat-engine effect of the Brownian ratchets with a time-dependent drive, in a system we call a Brownian pump. Reference~\cite{Raz2016} shows how to realize any nonequilibrium steady state using such time-dependent driving in systems called stochastic pumps. The difference is that they did not consider locality,  so their pumps are effectively few-body systems. Reference~\cite{Machado2023} shows how to realize generic CAs using time-dependent Hamiltonians of oscillators, with infinite onsite Hilbert space. We show how to do the same thing using finite classical onsite state spaces. For completeness, we provide an explicit implementation of this construction for Toom's rule in a subsequent subsection.

The pump consists of $2n$ copies of the system ($n$ leader copies and $n$ follower copies) evolving at temperature $T = 1/\beta$ under a time-dependent Hamiltonian with timescale $\tau$. We furthermore take $n$ to be odd to simplify the proof.
Like the ratchet, when the pump operates at finite temperatures and energy parameters, the simulation will be noisy. 
However, in the limit $1 \ll \tau \ll e^{\beta J}$ we will approach error-free (and synchronous) operation. 

\subsection{Definitions}

We again consider a collection of spins whose state is given by $\sigma = \{ \pm 1 \}^N$.
Using Def.~\ref{def:CA}, we can now define the synchronous dynamics of the PCA.
\begin{definition}[Synchronous PCA dynamics]
The synchronous dynamics corresponding to the PCA $A$, denoted $\mca_\sync$, is a discrete-time Markov process that updates all of the spins in parallel. Given $\sigma(t)$, the spin at site $i$ at the next time step is flipped according to
\begin{equation}
\Pr[\sigma_i(t+1)=-\sigma_i(t) \mid \sigma(t)] = p_i^A( \sigma(t)),
\end{equation}
where we have normalized the discrete clock rate to one.
\end{definition}
These are the dynamics that are most commonly meant by ``cellular automaton dynamics.''

In order to compare our pump dynamics to the underlying PCA, we define $\delta$-closeness for discrete-time Markov processes.
\begin{definition}[$\delta$-closeness]
Two local discrete-time Markov processes $\mathcal M$ and $\mathcal N$ are $\delta$-close if 
\begin{equation}
    |p^{\mathcal M}_i(\sigma) - p^\mathcal{N}_i(\sigma)| \le \delta
\end{equation}
for all locations $i$, states $a$, and configurations $\sigma$. 
\end{definition}

It will be useful to define the following function, which we call the majority tail. 
Given $n$ independent processes with probability $p$, the probability that more than half of them happen is given by the majority tail function 
\begin{equation}
B_n(p) = \sum_{k=m}^n \binom{n}{k} p^k (1-p)^{n-k} \le 2^np^m,
\end{equation}
where $m = (n+1)/2$ for odd $n$.

\subsection{Construction}

For simplicity we will first construct the Brownian pump for a deterministic CA $A$, so that $p^{ A}_i(\s_f) \in \{ 0, 1 \}$. The extension to PCAs will be given in Sec.~\ref{sub:pumppcas}. We again consider two flavors of spin on each vertex, called leader ($\ell$) and follower ($f$). Furthermore, we place a finite odd number $n$ of each type of spin (indexed by $\alpha$) on each vertex as in Sec.~\ref{sec:warmup}. We define the effective degrees of freedom
\begin{align}
\sigma_{\ell,i}(t) &= \sgn \left(\sum_\alpha \s_{\ell, i, \alpha}(t) \right), \nonumber\\
\sigma_{f,i}(t) &= \sgn \left(\sum_\alpha \s_{f, i, \alpha}(t) \right)
\end{align}
as the majority vote values of the leader or follower spins on each site, and denote their configurations as $\sigma_\ell$ and $\sigma_f$.

Like in the few-body pump, we will use two types of terms to build the Hamiltonian. The first will hold the follower spins fixed and update the leaders, while the second will hold the leader spins fixed and update the followers to match the leaders. The Hamiltonian is periodic with period $2\tau$, $H(t+2\tau)= H(t)$, and given by 
\begin{align}
	H_A(t) &= \begin{cases}
		H_1, & 0 \le  t  < \tau \\
		H_2, & \tau\le  t  < 2\tau,
	\end{cases}, \nonumber\\
	H_1 &= -J \sum_{i, \alpha < \beta} \sigma_{f, i, \alpha} \sigma_{f, i, \beta} - \sum_{i, \alpha} h_i(\sigma_{f}) \sigma_{\ell, i, \alpha}, \nonumber\\
	H_2 &= -J \sum_{i, \alpha < \beta} \sigma_{\ell, i, \alpha} \sigma_{\ell, i, \beta} - J \sum_{i, \alpha} \sigma_{f, i, \alpha} \sigma_{\ell, i, \alpha}, \label{eqn:generic_pump}
\end{align}
where the field 
\begin{equation}
h_i(\sigma_{f}) =  \begin{cases}
    \hphantom{-} J\s_{f, i}, & p_i^A(\sigma_f) = 0,\\
    -J\s_{f, i}, & p_i^A(\sigma_f) = 1,
\end{cases}
\end{equation}
encourages $\s_{\ell, i, \alpha}$ to relax to the correct value. 

During the first half of the cycle, the first term in $H_1$ ensures that the follower spins do not update. The other term attempts to evolve the leader spins forward in time according to $p_i^A(\sigma_{f})$. 
During the second half of the cycle, the first term in $H_2$ ensures that the leader spins do not update while the other term makes the followers match the leaders. 

We now define the Brownian pump dynamics as the discrete-time Markov process induced on the effective leader spins $\sigma_\ell$.
\begin{definition}[Brownian pump]
For any PCA $A$, the Brownian pump dynamics $\mcp_A(J, T, n, \tau)$ is a local discrete-time Markov process defined from the Glauber dynamics of the Floquet Hamiltonian $H_A(t)$ (Eq.~\ref{eqn:generic_pump} for deterministic CAs, Eq.~\ref{eqn:pump_PCAs} for generic PCAs) at temperature T and odd $n$.
The flip probabilities are given by
\begin{equation}
    p^{\mcp_A}_i(\sigma_\ell) \equiv \Pr[\sigma_{\ell, i}(t + k\tau) = -\sigma_{\ell, i}(t) | \sigma_f(t)],
\end{equation}
where 
\begin{equation}
\sigma_{\ell,i}(t) = \sgn \left(\sum_\alpha \s_{\ell, i, \alpha}(t) \right)
\end{equation}
are the majority vote values of the $n$ leader spins on each site at discrete times $t \in k\tau\mathbb{Z}$, where $k=2$ for discrete CAs and $k=3$ for probabilistic CAs.
\end{definition}
As in Sec.~\ref{sec:warmup}, this Floquet system approaches exact simulation of the CA as long as $n > r+1$, in the limit
\begin{equation}
	1 \ll \tau \ll e^{\beta J}, \label{eq:floq_limit}
\end{equation}
where time is measured in units of the expected time between thermal updates at a given site. Note that this implies $J \gg T$.
We can explain the two sides of the limit separately. Recall that we have $1 \ll \tau$ to make sure that we hold the Hamiltonian long enough for every leader spin to update or for every follower spin to match, and we have $\tau \ll e^{\beta J}$ so that we don't hold the Hamiltonian for so long that the spins become thermally excited.

\subsection{Brownian pump simulation}

Let us bound the error rates like in Sec.~\ref{sub:rates}. Importantly, the pump operates in an effectively discrete time, so we should calculate error probabilities per cycle instead of continuous error rates. 

\begin{theorem} [Brownian pumps simulate discrete-time CA dynamics]
    Let $A$ be a CA with neighborhood size $\size$, and $\mca_\sync$ its synchronous CA dynamics. For every $\delta>0$, there exists a Brownian pump $\mcp_A(J, T, n, \tau)$ that is $\delta$-close to $\mca_\sync$. The simulation closeness is set by
    \begin{align}
        \delta \le 2^n \big( (r p_f)^m &+ (r \tau R_\mathrm{err})^m \nonumber\\ 
        &+ (e^{-\tau R_\mathrm{update}} + \tau R_\mathrm{leak})^m\big),
    \end{align}
    where $m=(n+1)/2$ is the minimum number of spins required to flip the majority vote of the $n$ copies, with 
    \begin{equation}
        p_f  \le C_n e^{-2\beta J (n-1)} + e^{-2\beta J}
    \end{equation}
    where $C_n$ depends only on $n$, and 
    \begin{align}
        R_\mathrm{err} &= \frac{1}{1+e^{2(n-1-r)\beta J}}, \nonumber\\
        R_\mathrm{update} &= \frac{1}{1+e^{-2\beta J}}, \nonumber\\
        R_\mathrm{leak} &= \frac{1}{1+e^{2\beta J}},
    \end{align}
    so that $\delta \rightarrow 0$ in the limit $1 \ll \tau \ll e^{\beta J}$ and $n > r+1$.
    \label{thm:pump}
\end{theorem}
We again construct the proof via a union bound on the different error channels.

The object that we have to calculate is $p^{{\mathcal P}_A}_i(\sigma_\ell),$ where the configurations $\sigma_\ell$ are not the configurations of individual spins but rather the majorities of the leader spins. This means that $\sigma_{\ell, i}$ does not determine the states of the individual spins $\s_{\ell, i, \alpha}$ uniquely, but rather stochastically. At the beginning of a cycle, assuming that all spins have had a chance to thermalize with respect to $H_2$---which $1\ll\tau$ ensures is indeed the case---the probability that a randomly chosen leader spin does not match the simulated state is 
\begin{equation}
    p_\ell \equiv \Pr(\sigma_{\ell, i, \alpha } \ne \sigma_{\ell, i}) \le (2^n-2) e^{-2\beta J (n-1)},
\end{equation}
via a union bound on the product of the number of nonuniform states and the minimum energy penalty for a nonuniform state. Similarly, the probability that a follower spin does not match the simulated state is
\begin{equation}
    p_f \equiv \Pr(\sigma_{f, i, \alpha } \ne \sigma_{\ell, i}) \le (2^n-2) e^{-2\beta J (n-1)} + e^{-2\beta J},
\end{equation}
also from a union bound on $\sigma_{\ell, i, \alpha}\ne \sigma_{\ell, i}$ and $\sigma_{f, i, \alpha}\ne \sigma_{\ell, i, \alpha}$. 

The first type of error is a ``bad input'' error during the evolution of $H_1$, where $\sigma_{\ell, i, \alpha}$ updates to the wrong state because one of the $\sigma_{f, j, \alpha}$ were in the wrong state to begin. For an individual $\alpha$, this error probability is $\le r p_f$ by a union bound. This only translates into a simulation error if at least half of the leader spins at site $i$ fail, so the total error is given by
\begin{equation}
    B_n(r p_f) \le 2^n (r p_f)^m.
\end{equation}

Then, the next type of error (still during $H_1$) is that one of the follower spins $\sigma_{f, i, \alpha}$ might update when it is not supposed to. As a worst-case scenario, say that all of the nearby leader spins are in a state such that $\sigma_{f, i, \alpha}$ wants to flip. At the same time, with probability $1-p_f$, all $n-1$ other $\sigma_{f, i}$ spins discourage it from flipping. The net rate is
\begin{equation}
    R_\text{err} = \frac{1}{1+e^{2(n-1-r)\beta J}},
\end{equation}
where $\size$ is the number of spins within the interaction range of $\mca$. The total error probability (assuming all followers match) is
\begin{equation}
    p_\textrm{err} = 1-e^{-\tau R_\text{err}} \le \tau R_\text{err},
\end{equation}
which again gets multiplied by $\size$ to account (via a union bound) for all of the follower spins that a single leader spin sees. Translating again into the probability that a majority of follower spins fail, we have
\begin{equation}
B_n(r \tau R_\text{err}) \le 2^n (r \tau R_\text{err})^m.
\end{equation}

Conditioning on $\sigma_{\ell, i, \alpha}$ seeing the correct local environment, if it needs to flip it does so at rate
\begin{equation}
R_\text{update} = \frac{1}{1+e^{-2\beta J}},
\end{equation}
because the flip lowers the energy by $2J$. The probability that an individual spin fails to update during this time is 
\begin{equation}
p_\text{update} \le e^{-\tau R_\text{update}}.
\end{equation}
Furthermore, once a spin has flipped it leaks back into the other state at rate
\begin{equation}
R_\text{leak} = \frac{1}{1+e^{2\beta J}},
\end{equation}
so the probability is 
\begin{equation}
B_n( e^{-\tau R_\text{update}} +\tau R_\text{leak}) \le 2^n (e^{-\tau R_\text{update}} + \tau R_\text{leak})^m
\end{equation}
for a majority of the leader spins to not update to the correct state. 

Combining these error probabilities, we arrive at Thm.~\ref{thm:pump}.

\subsection{Intrinsically probabilistic CAs} \label{sub:pumppcas}

We now generalize the proof of Thm.~\ref{thm:pump} to PCAs. 
The Hamiltonian is periodic with period $3\tau$, $H(t+3\tau)= H(t)$, and given by 
\begin{align}
	H_A(t) &= \begin{cases}
		H_1, & 0 \le  t  < \tau \\
		H_2, & \tau\le  t  < 2\tau \\
		H_3, & 2\tau\le  t  < 3\tau,
	\end{cases} \nonumber\\
	H_1 &= -J \sum_{i, \alpha < \beta} \sigma_{f, i, \alpha} \sigma_{f, i, \beta} - \sum_{i, \alpha} \sigma_{\ell, i, \alpha} h_i(\sigma_{f}), \nonumber\\
    H_2 &= -J \sum_{i, \alpha < \beta} \sigma_{\ell, i, \alpha} \sigma_{\ell, i, \beta}, \nonumber\\
    H_3 &= -J \sum_{i, \alpha < \beta} \sigma_{\ell, i, \alpha} \sigma_{\ell, i, \beta} - J \sum_{i, \alpha} \sigma_{f, i, \alpha} \sigma_{\ell, i, \alpha}, \label{eqn:pump_PCAs}
\end{align}
where the field
\begin{equation}
h_i(\sigma_{f}) =  \begin{cases}
    \hphantom{-} J\s_{f, i}, & p_i^A(\sigma_f) = 0,\\
    -J\s_{f, i}, & p_i^A(\sigma_f) = 1, \\
    h_i'(\s_f), & \text{otherwise}
\end{cases}    
\end{equation}
is chosen to ensure the correct update rate for $\sigma_{\ell, i}$.
During $H_2$, the follower spins are completely free to update randomly, and the only purpose of this step is to allow the leader spins to perform their majority vote. 

We have already bounded the error probability for the first two cases. Here, we choose the value of $h'_i(\s_f)$ such that the probability that a majority of the leader spins flip during an entire cycle is $p_i^A(\sigma_{f, \alpha})$. First, define the probability for a single leader spin $\s_{\ell, i, \alpha}$ to flip after $H_1$ as 
\begin{equation}
    q_i(\s_f) = \Pr(\sigma_{\ell, i, \alpha} = -\s_{f, i, \alpha} | \sigma_f).
\end{equation}
After $H_1$, the expectation of $\sigma_{\ell, i, \alpha}$ is 
\begin{align}
    \langle \sigma_{\ell, i, \alpha} \rangle  &= \sigma_{f, i, \alpha} [1 - 2 q_i(\sigma_f)] \nonumber\\
    &= \tanh (\beta h'_i(\sigma_f)),
\end{align}
so $h'_i(\s_f)$ is given by
\begin{equation}
    h'_i(\s_f) = T \tanh^{-1} (\sigma_{f, i, \alpha} [1 - 2 q_i(\sigma_f)]),
\end{equation}
and we just have to choose $q_i(\s_f)$ given $p_i^A(\sigma_f)$.

To do so, we want to calculate $p_\text{flip}(\s_f)$, the probability that $\s_{\ell, i}$ flips during $H_1$ and $H_2$. Define
\begin{equation}
    P_k = \Pr(\text{majority vote flips} | \text{$k$ spins flip})
\end{equation}
as the probability that the majority vote of the leader spins flips, given that $k$ individual spins flip. 
Then, the full flip probability is given by 
\begin{equation}
    F_n(q) = \sum_{k=m}^n \binom{n}{k} q^k (1-q)^{n-k}P_k,
\end{equation}
which differs from $B_n(q)$ by the factor of $P_k$. We want to choose 
\begin{equation}
q_i(\sigma_f) = F^{-1}_n(p_i^A(\sigma_f)),
\end{equation}
to arrive at 
\begin{equation}
\Pr[\sigma_{\ell, i}(t + 3\tau) = -\sigma_{\ell, i}(t) | \sigma_f(t)] = p^{A}_i(\sigma_\ell).
\end{equation}
Combined with the bounds on errors from the previous subsection, this proves Thm.~\ref{thm:pump} for generic PCAs.

\subsection{A pump for Toom's rule}

\begin{figure}
	\centering
	\includegraphics[width=\linewidth]{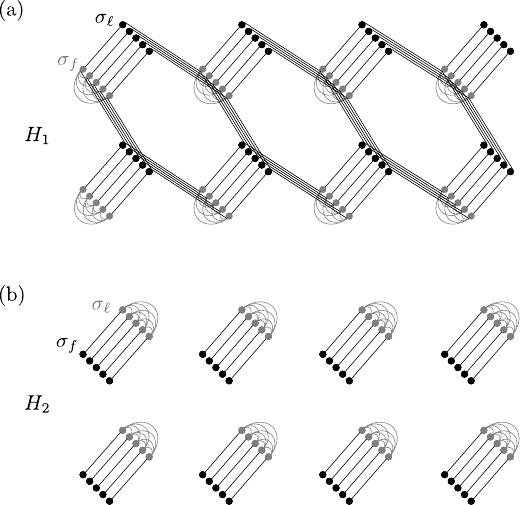}
	\caption{Brownian pump for Toom's rule, with $n=5$. (a) Under $H_1$, the follower spins (in gray) are held fixed while the leader spins (in black) update via the majority vote in Toom's rule. (b) Under $H_2$, the leader spins (in gray) are held fixed while the follower spins (in black) update to match the leaders.}
	\label{fig:toom_pump}
\end{figure}

Finally, let us illustrate the pump construction for Toom's rule.
We will still use $n$ leader and follower spins per site, with $n=5$ in Fig.~\ref{fig:toom_pump}. We still use onsite couplings to hold one of the spin types fixed while updating the other, but the leader update term can now be a bit simpler.

The Hamiltonian is now periodic with period $2\tau$, $H(t+2\tau)= H(t)$, and given by 
\begin{equation}
	H_A(t) = \begin{cases}
		H_1, & 0 \le  t  < \tau \\
		H_2, & \tau\le  t  < 2\tau,
	\end{cases} \label{eqn:toom_pump}
\end{equation}
where now 
\begin{align}
	H_1 = &-J \sum_{i, \alpha < \beta} \sigma_{f, i, \alpha} \sigma_{f, i, \beta} \nonumber\\
    &- J \sum_{i, \alpha} \sigma_{\ell, i, \alpha} (\sigma_{f, i, \alpha} + \sigma_{f, i+\hat{x}, \alpha} + \sigma_{f, i+\hat{y}, \alpha}) 
\end{align}
and
\begin{equation}
	H_2 = -J \sum_{i, \alpha < \beta} \sigma_{\ell, i, \alpha} \sigma_{\ell, i, \beta} - J \sum_{i, \alpha} \sigma_{f, i, \alpha} \sigma_{\ell, i, \alpha},
\end{equation}
so that all couplings are ferromagnetic.
For an illustration of the Hamiltonian schedule, see Fig.~\ref{fig:toom_pump}.

During the first half of the cycle, $H_1$ ensures that the follower spins do not update and implements the Toom update for the leader spins. During the second half of the cycle, $H_2$ ensures that the leader spins do not update and attempts to make the follower spins match the leaders.  This Floquet system exactly simulates the CA in the limit
\begin{equation}
	1 \ll \tau \ll e^{\beta J}, 
\end{equation}
just as in the generic case.

\section{Discussion and Outlook} \label{sec:discussion}

In this work, we have given universal constructions for simulating generic active dynamics with simple nonequilibrium drives.
{\it Brownian ratchets} use static Hamiltonians coupled to heat baths at different temperatures to simulate continuous-time dynamics. {\it Brownian pumps} use periodic time-dependent Hamiltonians to simulate discrete-time  dynamics. As a concrete example, we constructed a Toom ratchet in which a maintained temperature difference preserves memory even under a symmetry-breaking field, something no equilibrium Ising model can do. These constructions raise a number of natural questions, a subset of which we discuss below. 

The first concerns the thermodynamic cost of autonomous error correction.
In the Brownian Toom ratchet, heat flows from the hot layer to the cold layer in the steady state. As in an ordinary ratchet, this heat current is the resource that drives the useful dynamics; here the useful task is not lifting a load, but sustaining the active erosion process that protects the memory. The natural cost is therefore the entropy production rate associated with the heat flow, set by the current and the temperature difference between the baths. 
The Brownian ratchet setting allows us to compare to standard results on entropy production from stochastic thermodynamics~\cite{Seifert2012, Parrondo2015, Mandal2017, Nardini2017, Dabelow2019, Fodor2022}.
Furthermore, this motivates the question: How much entropy production is necessary to protect a $n$ bits of information for time $\tau_\text{mem}$?

Another question is how faithfully the ratchet realizes the target active dynamics away from the ideal simulation limit. We have proven that the Toom ratchet approaches Toom-like dynamics in a controlled regime, but our numerics show stability even when the clock rates are equal and the energy scales are not strongly separated. In this intermediate regime, the model is not literally Toom's rule, but it still achieves the same qualitative task of protecting memory against a bias. More generally, one can ask when a ratchet or pump lies in the same coarse-grained universality class as the PCA it was built to simulate.

In another direction, it is interesting to speculate as to whether any of our constructions can be made technologically useful beyond proof-of-principle models. As magnetic memory devices shrink, domain density increases, raising the risk of stray effective fields from neighboring domains causing unwanted bit flips. Robustness to strong external fields is also essential in satellite computing, where onboard memory must withstand the harsh magnetic environment of space. It is theoretically encouraging that we can realize robust error-correcting dynamics like Toom's rule using simple autonomous drives, but can these dynamics lead to technologically relevant driven memories?

While we considered two particular forms of driving (time dependence and temperature gradients), many other forms of driving exist. With varying degrees of separation from our constructions, we could consider a gradient in the chemical potential for a conserved charge, 
a fixed Hamiltonian at a time-dependent temperature, or sliding multiple systems across each other at constant velocity. All of these mechanisms keep a system out of equilibrium and lead to interesting transport effects~\cite{Reimann2002BrownianMotors}. Are all of these mechanisms universal for either discrete- or continuous-time PCA dynamics?

It is also natural to ask which classes of active dynamics can simulate which others. The basic distinction between our pumps and ratchets is synchronization: pumps come with a global clock, while ratchets run in continuous time. In one direction, discrete-time dynamics can approximate aspects of continuous-time dynamics, for example through checkerboard Glauber updates that alternate between sublattices and prevent neighboring spins from updating simultaneously~\cite{Heermann1990}. Using more sublattices makes this approximation more faithful, because fewer nearby degrees of freedom are updated in the same step. The more interesting direction is the reverse: can a continuous-time autonomous system robustly generate the synchronization needed to simulate a discrete-time PCA~\cite{cook2008self}? If not, synchronization itself may be an additional nonequilibrium resource, separating asynchronous active dynamics from clocked dynamics that require some form of global coordination.

It would also be interesting to extend these ideas to quantum systems. In the few-body setting, quantum heat engines and refrigerators have been studied extensively~\cite{Scovil1959, Kosloff2014, Mitchison2019, Aamir2025}, in connection with the broader fields of quantum thermodynamics~\cite{Vinjanampathy2016, Goold2016, Millen2016, Strasberg2017, Gour2015, YungerHalpern2016} and active quantum matter~\cite{Verstraete2009, Adachi2022, Yamagishi2024, Khasseh2025, Guo2025}. In the many-body setting, the natural question is whether analogous ratchet and pump mechanisms can realize genuinely quantum active dynamics or phases. In the classical constructions above, the Hamiltonian sets energy barriers and transition rates for thermal updates. In a quantum system, the same Hamiltonian also generates coherent real-time evolution, while the reservoirs needed to drive the dynamics can cause decoherence. Understanding when coherent evolution helps, when dissipation washes it out, and what genuinely quantum information or order can survive in such a driven dissipative setting is an open question.

A concrete motivation comes from quantum error correction.
A long-term goal in the field has been to find local mechanisms for protecting quantum information.  While some of this literature focuses on ``self-correcting'' memories, in which passive error correction by coupling to a cold bath is sufficient to protect a quantum memory, other works attempt to build local active decoders, motivated by local classical error correction. The seminal work by G\'acs~\cite{Gacs2001} claims that local classical error correction is possible in 1D, and motivates analogous quantum decoders in 2D~\cite{harrington2004analysis, Breuckmann2017LocalDecoders}, see also~\cite{Duennweber2026}. More recently, an older 1D classical decoder by Tsirelson~\cite{Tsirelson} has motivated another 2D quantum decoder~\cite{Balasubramanian2024}. While the decoders in those works are unavoidably complicated, the current work should explain some of the excitement about local decoding. Our constructions suggest a possible route from such decoding rules to autonomous dynamics: once a local decoder is written as PCA dynamics on the syndrome information, no site-resolved measurement or feedback is required. Instead, the decoding scheme can be translated into a simple time-dependent Hamiltonian coupled to a cold bath or, if the scheme is time-independent and asynchronous, a fixed Hamiltonian coupled to two heat baths. 

More broadly, many-body Brownian ratchets provide a static-Hamiltonian setting for active matter. Historically, Brownian ratchets have been studied primarily as mechanisms for transport and rectification. Our results show that, in many-body systems, the same basic thermodynamic ingredient can do more: a steady heat current can power local information processing and stabilize robust nonequilibrium behavior. This suggests that Brownian ratchets may serve as a useful DC counterpart to periodically driven systems, thus furnishing a new playground for exploring nonequilibrium many-body phases stabilized by thermodynamic currents.

\section*{Acknowledgments}

We thank Simon Blanch, Sarang Gopalakrishnan, David Huse, Lauren Li, Yaodong Li, David Long, Nicholas O'Dea, Tibor Rakovszky, and Mike Zaletel for helpful comments. C.S. thanks Oliver Hart, Aaron J Friedman, and Jack H. Farrell for inspiring conversations. C.S. and V.K. are supported by the Office of Naval
Research Young Investigator Program (ONR YIP) under Award Number N000142412098. V.K. also acknowledges support from the Packard Foundation through a
Packard Fellowship in Science and Engineering. E.L. was supported by a Miller Research Fellowship. 

\bibliography{refs}
	
\end{document}